\documentclass{amc}
 \usepackage{amsmath}
 \usepackage{amssymb}
 \usepackage{epsfig} 
 \usepackage{graphics} 

\def\currentvolume{X}
 \def\currentissue{X}
  \def\currentyear{200X}
   \def\currentmonth{XX}
    \def\ppages{X--XX}
\usepackage[pdfpagemode=None]{hyperref}

\newcommand{\bi}{\begin{itemize}}
\newcommand{\ei}{\end{itemize}}
\newcommand{\ben}{\begin{enumerate}}
\newcommand{\een}{\end{enumerate}}
\newcommand{\beq}{\begin{equation}}
\newcommand{\eeq}{\end{equation}}
\newcommand{\beqa}{\begin{eqnarray}}
\newcommand{\eeqa}{\end{eqnarray}}

\newcommand{\circr}{\mbox{\textcircled{\scriptsize $R$}}}

\newcommand{\circz}{\mbox{\textcircled{\scriptsize $Z$}}}

\newcommand{\circzb}{\mbox{\textcircled{\tiny $Z_B$}}}

\newcommand{\circzm}{\mbox{\textcircled{\tiny $Z_M$}}}

\newtheorem{theorem}{Theorem}

\newtheorem{lemma}{Lemma}

\theoremstyle{definition}
\newtheorem{definition}{Definition}

\newtheorem{example}{Example}

\title[Zig-zag and Replacement Product Graphs and LDPC
Codes]{Zig-zag and Replacement Product Graphs and LDPC Codes}

\author[Christine A. Kelley \and Deepak Sridhara \and Joachim Rosenthal]{}


\email{ckelley@math.ohio-state.edu; deepak.sridhara@seagate.com;
rosen@math.unizh.ch}

\thanks{This work was supported in part by the NSF Grant
  No. CCR-ITR-02-05310 and the Swiss NSF Grant No. 113251, and
  conducted in part when the first author was at the Fields
  Institute in Toronto, Canada, and when the second and third
  authors were at the University of Z\"urich.
  }

\begin{document}
\maketitle


\centerline{\begin{tabular}{ccc} \vspace{-0.00in}Christine A.
    Kelley&\vspace{-0.00in}Deepak Sridhara&\vspace{-0.00in}
    Joachim Rosenthal\\
    \vspace{-0.00in}The Ohio State University& \vspace{-0.00in}
    1251 Waterfront Place &
    \vspace{-0.00in} Institut f$\ddot{\mbox{u}}$r Mathematik\\
    \vspace{-0.00in}Department of Mathematics& \vspace{-0.00in}
    Seagate Technology&
    \vspace{-0.00in}Universit$\ddot{\mbox{a}}$t Z$\ddot{\mbox{u}}$rich \\
    \vspace{-0.00in}Columbus, OH 43210& \vspace{-0.00in}
    Pittsburgh, PA 15222 & \vspace{-0.00in} CH-8057,
    Switzerland.\\
    \vspace{-0.00in} {\tt ckelley}@math.ohio-state.edu&
    \vspace{-0.00in} {\tt deepak.sridhara}@seagate.com&
    \vspace{-0.00in} {\tt rosen}@math.unizh.ch
\end{tabular}
}


\medskip

 \centerline{(Communicated by Aim Sciences)}
 \medskip

\begin{abstract}
  The performance of codes defined from graphs depends on the
  expansion property of the underlying graph in a crucial way.
  Graph products, such as the zig-zag product~\cite{re02} and
  replacement product provide new infinite families of constant
  degree expander graphs. The paper investigates the use of
  zig-zag and replacement product graphs for the construction of
  codes on graphs~\cite{ta81a}.  A modification of the zig-zag
  product is also introduced, which can operate on two unbalanced
  biregular bipartite graphs.

\end{abstract}

\section{Introduction}              \label{sec-1}
Expander graphs are of fundamental interest in mathematics and
engineering and have several applications in computer science,
complexity theory, designing communication networks, and coding
theory~\cite{li03r,al86,ta84}.  In a remarkable paper~\cite{re02}
Reingold, Vadhan, and Wigderson introduced an iterative construction
which leads to infinite families of constant degree expander graphs.
The iterative construction is based on the {\em zig-zag graph
product} introduced by the authors in the same paper. The zig-zag
product of two regular graphs is a new graph whose degree is equal
to the square of the degree of the second graph and whose expansion
property depends on the expansion properties of the two component
graphs. In particular, if both component graphs are good expanders,
then their zig-zag product is a good expander as well. Similar
things can be said about the replacement product.

Since the work of Sipser and Spielman~\cite{si96} it has been
well known that the performance of codes defined on graphs
depends on the expansion property of the underlying graph in a
crucial way. Several authors have provided constructions of codes
from graphs whose underlying graphs are good expanders.  In
general, a graph that is a good expander is particularly suited
for the message-passing decoder that is used to decode low
density parity check (LDPC) codes, in that it allows for messages
to be dispersed to all nodes in the graph as quickly as possible.
Furthermore, graphs with good expansion yield LDPC codes with
good minimum distance and pseudocodeword
weights~\cite{ja03,ke06u1,la00p,ro00p,si96}.

Probably the most prominent example of expander graphs are the class
of {\rm Ramanujan} graphs which are characterized by the property
that the second eigenvalue of the adjacency matrix is minimal inside
the class of $k$-regular graphs on $n$ vertices. This family of
`maximal expander graphs' was independently constructed by Lubotzky,
Phillips and Sarnak~\cite{lu88a} and by Margulis~\cite{ma88a2}. The
description of these graphs and their analysis rely on deep results
from mathematics using tools from graph theory, number theory, and
representation theory of groups~\cite{lu94}.  Codes from Ramanujan
graphs were constructed and studied by several
authors~\cite{la00p,ro00p,ti97}.

Ramanujan graphs have the drawback that they exist only for a
limited set of parameters. In contrast, the zig-zag product and the
replacement product can be performed on a large variety of component
graphs. The iterative construction also has a lot of engineering
appeal as it allows one to construct larger graphs from smaller
graphs as one desires. This was the starting point of our research
reported in~\cite{ke03p}.

In this paper we examine the expansion properties of the zig-zag
product and the replacement product in relation to the design of
LDPC codes. We also introduce variants of the zig-zag scheme that
allow for the component graphs to be unbalanced bipartite graphs.
In our code construction, the vertices of the product graph are
interpreted as sub-code constraints of a suitable linear block
code and the edges are interpreted as the code bits of the LDPC
code, as originally suggested by Tanner in~\cite{ta81a}. Codes
obtained in this way will be referred to as {\em generalized
  LDPC} (GLDPC) {\em codes}. By choosing component graphs with
relatively small degree, we obtain product graphs that are
relatively sparse.  Examples of each product and resulting LDPC
codes are given to illustrate the results of this paper. Some of
the examples use Cayley graphs as components, and the resulting
product graph is also a Cayley graph with the underlying group
being the semi-direct product of the component groups, and the
new generating set being a function of the generating sets of the
components \cite{al01p}. Simulation results reveal that LDPC
codes based on zig-zag and replacement product graphs perform
comparably to, if not better than, random LDPC codes of
comparable block lengths and rate. The vertices of the product
graph must be fortified with strong (i.e., good minimum distance)
sub-code constraints, in order to achieve good performance with
message-passing (or, iterative) decoding.

The paper is organized as follows. Section~\ref{sec-2} discusses
preliminaries on the formal definition of expansion for a
$d$-regular graph and the best one can achieve in terms of
expansion. Furthermore, expansion for a general graph is
discussed. Section~\ref{sec-3} describes the original zig-zag
product and replacement product. A first result on the girth and
diameter of the replacement product is derived.

Section~\ref{sec-4} contains a new zig-zag product construction of
unbalanced bipartite graphs. The main result is
Theorem~\ref{zztheorem} which essentially states that the
constructed bipartite graph is a good expander graph if the two
component graphs are good expanders.

Section~\ref{sec-5} is concerned with applications to coding
theory. The section contains several code constructions using the
original and the unbalanced bipartite zig-zag products and the
replacement product.  Simulation results of the LDPC codes
constructed in Section~\ref{sec-5} are presented in
Section~\ref{sec-6}.  Section~\ref{sec-7} introduces a new
iterative construction for an unbalanced bipartite zig-zag
product and the replacement product to generate families of
expanders with constant degree. For completion, the iterative
construction for the original zig-zag product from \cite{re02} is
also described.  The expansion for the iterative families are
also discussed.  Section~\ref{sec-8} summarizes the results and
concludes the paper.  \vspace{0.1in}

\section{Preliminaries}                                \label{sec-2}
In this section, we review the basic graph theory notions used in
this paper.

Let $G= (V,E)$ be a graph with vertex set $V$ and edge set $E$.  The
number of edges involved in a path or cycle in $G$ is called the
{\em length} of the path or cycle. The   {\em girth} of $G$ is the
length of the shortest cycle in $G$. If $x,y \in V$ are two vertices
in $G$, then the {\em distance} from $x$ to $y$ is defined to be the
length of the shortest path from $x$ to $y$. If no such path exists
from $x$ to $y$, then we say the distance from $x$ to $y$ is
infinity. The {\em diameter} of $G$ is the maximum distance among
all pairs of vertices of $G$.

Intuitively, a graph has good expansion if any small enough set of
vertices in the graph has a large enough set of vertices connected
to it. It is now almost common knowledge that for a graph to be a
good expander~\cite{si96}, the second largest eigenvalue of the
adjacency matrix $A$ must be as small as possible compared to the
largest eigenvalue~\cite{ta84}.  For a $d$-regular graph $G$, the
index (or, the largest eigenvalue) of the adjacency matrix $A$ is
$d$. Hence, by {\em normalizing} the entries of $A$ by the factor
$d$, the normalized matrix $\tilde{A}=\frac{1}{d}A$ has the largest
eigenvalue equal to~$1$. \vspace{0.1in}

\begin{definition}
  {\rm Let $G$ be a $d$-regular graph on $N$ vertices. Denote by
    $\lambda(G)$ the second largest eigenvalue of the normalized
    adjacency matrix $\tilde{A}$ representing $G$. $G$ is said to
    be a {\em $(N,d,\lambda)$-graph } if $\lambda=\lambda(G)$.  }
\end{definition}
\vspace{0.1in}


 In this paper, we will follow the definition provided
in~\cite{al01p,me03} for a graph to be an {\em expander}.
\vspace{0.1in}

\begin{definition} {\rm A sequence of
    graphs is said to be an {\em expander family} if for every
    (connected) graph $G$ in the family, the second largest
    eigenvalue $\lambda(G)$ is bounded below some constant
    $\kappa < 1$. In other words, there is an $\epsilon >0$ such
    that for every graph $G$ in the family,
    $\lambda(G)<1-\epsilon$. A graph belonging to an expander
    family is called an {\em expander graph}.}
\end{definition}
\vspace{0.1in}

Alon and Boppana have shown that for a $d$-regular graph $G$, as the
number of vertices $n$ in $G$ tends to infinity, $\lambda(G)
\ge\frac{2\sqrt{d-1}}{d}$~\cite{al86}.
 For $d$-regular (connected) graphs, the best possible expansion
based on the eigenvalue bound is achieved by Ramanujan graphs that
have $\lambda(G) \le \frac{2\sqrt{d-1}}{d}$~\cite{lu88a}.  Hence,
Ramanujan graphs are optimal in terms of the eigenvalue gap
$1-\lambda(G)$.

The definition of expansion to $d$-regular graphs can be similarly
extended to $(c,d)$-regular bipartite graphs as defined below and
also to general irregular graphs. \vspace{0.1in}

\begin{definition}{\rm
A graph $G=(X,Y;E)$ is {\em $(c,d)$-regular bipartite} if the set of
vertices in $G$ can be partitioned into two disjoint sets $X$ and
$Y$ such that all vertices in $X$ (called {\em left} vertices) have
degree $c$ and all vertices in $Y$ (called {\em right} vertices)
have degree $d$ and each edge  $e\in E$ of $G$ is incident with one
vertex in $X$ and one vertex in $Y$, i.e., $e=(x,y), x\in X, y\in
Y$.}
\end{definition}
\vspace{0.1in}

\begin{definition}
{\rm A $(c,d)$-regular bipartite graph $G$ on $N$ left vertices  and
$M$ right vertices is said to be a {\em $(N,M,c,d,\lambda)$-graph}
if the second largest eigenvalue of the normalized adjacency matrix
$\tilde{A}$ representing $G$ is $\lambda$. }
\end{definition}
\vspace{0.1in}

The largest eigenvalue of a $(c,d)$-regular graph is $\sqrt{cd}$.
Once again, normalizing the adjacency matrix of a $(c,d)$-regular
bipartite graph by its largest eigenvalue $\sqrt{cd}$, we have that
the (connected) graph is a good expander if second largest
eigenvalue of its normalized adjacency matrix is bounded away from 1
and is as small as possible.

To normalize the entries of an irregular graph $G$ defined by the
adjacency matrix $A=(a_{ij})$, we scale each $(i,j)^{th}$ entry in
$A$ by $\frac{1}{r_ic_j}$, where $r_i$ and $c_j$ are the $i^{th}$
row weight and $j^{th}$ column weight, respectively, in $A$. It is
easy to show that the resulting normalized adjacency matrix has its
largest eigenvalue equal to one. The definition of an expander for
an irregular graph $G$ can be defined analogously. \vspace{0.1in}

\section{Graph Products}    \label{sec-graph}             \label{sec-3}

In designing codes over graphs,  graphs with good expansion,
relatively small degree, small diameter, and large girth are
desired. Product graphs give a nice avenue for code construction, in
that taking the product of small graphs suitable for coding can
yield larger graphs (and therefore, codes) that preserve these
desired properties. Standard graph products, however, such as the
Cartesian product, tensor product, lexicographic product, and strong
product, all yield graphs with large degrees. Although sparsity is
not as essential for generalized LDPC codes, large degrees
significantly increase the complexity of the decoder.


In this section we describe the zig-zag product
of~\cite{li03r,re02}, introduce a variation of the zig-zag product
that holds for bi-regular (unbalanced) bipartite graphs, and review
the replacement product. In each case, the expansion of the product
graph with respect to the expansion of the component graphs is
examined. When the graph is regular-bipartite, this bi-regular
product yields the product in~\cite{li03r,re02}.  In addition to
preserving expansion, these products are notable in that the
resulting product graphs have degrees dependent on only one of the
component graphs, and therefore can be chosen to yield graphs
suitable for coding.







\vspace{0.1in}

\subsection{Zig-zag product}
Let $G_1$ be a $(N_1,d_1,\lambda^{(1)})$-graph and let $G_2$ be a
$(d_1,d_2,$ $\lambda^{(2)})$-graph. Randomly number the edges
around each vertex of $G_1$ by $\{1,\ldots,d_1\}$, and each
vertex of $G_2$ by $\{1,\ldots,d_1\}$.Then the zig-zag product $G
= G_1 \circz G_2$ of $G_1$ and $G_2$, as introduced
in~\cite{li03r,re02}, is a $(N_1\cdot d_1,d_2^2,\lambda)$-graph
defined as follows\footnote{ This is actually the second
  presentation of the zig-zag product given in \cite{re02}; the
  original description required $\ell = k[i]$ in step 2 of the
  product, i.e. each endpoint of an edge had to have the same
  label.  }:

\begin{itemize}
\item vertices of $G$ are represented as ordered pairs $(v,k)$,
  where $v\in \{1,2,\ldots,N_1\}$ and $k\in\{1,2,\dots,d_1\}$.
  That is, every vertex in $G_1$ is replaced by a cloud of
  vertices of $G_2$.
\item edges of $G$ are formed by making two steps on the small
  graph and one step on the big graph as follows:
\begin{itemize} \item a step ``{\em zig}'' on the small graph
  $G_2$ is made from vertex $(v,k)$ to vertex $(v,k[i])$, where
  $k[i]$ denotes the $i^{th}$ neighbor of $k$ in $G_2$, for $i\in
  \{1,2,\dots,d_2\}$.
\item a deterministic step on the large graph $G_1$ is made from
  vertex $(v,k[i])$ to vertex $(v[k[i]],\ell)$, where $v[k[i]]$
  is the $k[i]^{th}$ neighbor of $v$ in $G_1$ and
  correspondingly, $v$ is the $\ell^{th}$ neighbor of $v[k[i]]$
  in $G_1$.
\item a final step ``{\em zag}'' on the small graph $G_2$ is made
  from vertex $(v[k[i]],\ell)$ to vertex $(v[k[i]],\ell[j])$,
  where $\ell[j]$ is the $j^{th}$ neighbor of $\ell$ in $G_2$,
  for $j\in \{1,2,\dots,d_2\}$.
\end{itemize}
Therefore, there is an edge between vertices $(v,k)$ and $(v[k[i]],
\ell[j])$ for $i,j \in \{1,\dots,d_2\}$.
\end{itemize}
\vspace{0.1in}

\begin{figure}[h]
  \centering{\resizebox{5in}{4in}{\includegraphics{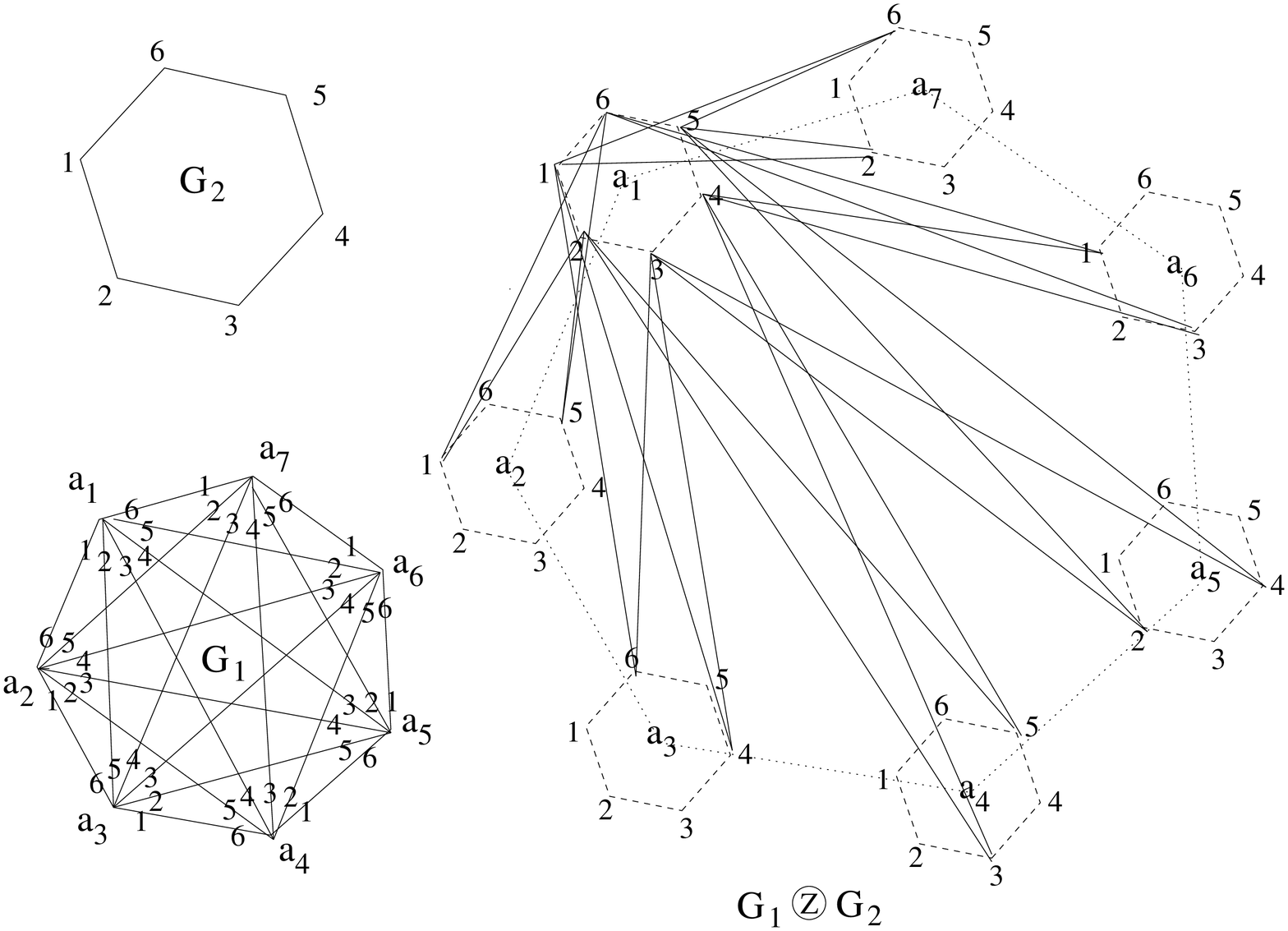}}}
\caption{Zig-Zag product of two graphs.} \label{zigzag_prod}
\end{figure}

\begin{example} Consider the zig-zag product graph $G = G_1 \circz G_2$
  depicted in Figure 1.  The edge from $(1,3)$ (``cloud $a_1$,
  vertex 3") to $(5,2)$ (``cloud $a_5$, vertex 2") is obtained by
  the following 3 steps:
\[ (1,3) \rightarrow (1,4) \rightarrow (5,3) \rightarrow (5,2) \]
The first step from 3 to 4 in cloud $a_1$ takes $(1,3)$ to
$(1,4)$.  The second step from $a_1$ to $a_5$ in $G_1$ takes
$(1,4)$ to $(5,3)$, since $a_5$ is the $4$th neighbor of $a_1$
and $a_1$ is the $3$rd neighbor of $a_5$ in the labeling around
the vertices of $G_1$. The final step in cloud $a_5$ takes
$(5,3)$ to $(5,2)$.  Similarly, vertex $(1,3)$ also connects to
$(5,4)$,$(3,4)$, and $(3,6)$ by these steps:
\[ (1,3) \rightarrow (1,4) \rightarrow (5,3) \rightarrow (5,4) \]
\[ (1,3) \rightarrow (1,2) \rightarrow (3,5) \rightarrow (3,4) \]
\[ (1,3) \rightarrow (1,2) \rightarrow (3,5) \rightarrow (3,6), \]
so the degree of vertex $(1,3)$ is $2^2 = 4$ as expected.
\end{example}

\vspace{0.2in}

It is shown in~\cite{re02} that the zig-zag product graph
$G=G_1\circz G_2$ is a $(N_1\cdot d_1,d_2^2,\lambda)$-graph with
$\lambda < \lambda^{(1)}+\lambda^{(2)}+[\lambda^{(2)}]^2$, and
further, that $\lambda <1$ if $\lambda^{(1)} <1$ and
$\lambda^{(2)}<1$. Therefore, the degree of the zig-zag product
graph depends only on the smaller component graph whereas the
expansion property depends on the expansion of both the component
graphs, i.e., it is a good expander if the two component graphs
are good expanders.

\vspace{0.1in}

\begin{lemma} Let $G_1$ and $G_2$ have girth $g_1$ and
  $g_2$, respectively. Then the zig-zag product graph
$G=G_1\circz G_2$ has girth $g=4$. 
\label{zz_lemma}
\end{lemma}
\vspace{0.1in}

\begin{proof}
  We show that any pair of vertices at distance two in $G_2$ are
  involved in a 4-cycle in $G$. Consider two vertices $(v_1,k_1)$
  and $(v_1,k_2)$ in the same cloud of $G$ that lie at distance
  two apart in $G_2$. Let $(v_1,k_3)$ be their common neighbor.
  In step 1 of the zig-zag product, an edge will start from
  $(v_1,k_1)$ and $(v_1,k_2)$ to $(v_1,k_3)$. Note that the
  deterministic step will then continue the edge from $(v_1,k_3)$
  to a specified vertex $(\tilde{v},\tilde{k})$ in another cloud.
  Therefore, with step 3, the actual edges in $G$ will go from
  $(v_1,k_1)$ to the neighbors of $(\tilde{v},\tilde{k})$, and
  from $(v_1,k_2)$ to the neighbors of $(\tilde{v},\tilde{k})$.
  Therefore, $(v_1,k_1)$ and $(v_1,k_2)$ are involved in a
  4-cycle provided $(\tilde{v},\tilde{k})$ does not have degree
  1. Since it is assumed $G_2$ is a connected graph with more
  than 2 vertices, there is a pair of vertices such that the
  resulting $(\tilde{v},\tilde{k})$ has degree $> 1$ in $G_2$.
\end{proof}
\vspace{0.1in}

We now consider the case when the two component graphs are Cayley
graphs~\cite{ro00p}. Suppose $G_1 = C(G_a,S_a)$ is the Cayley
graph formed from the group $G_a$ with $S_a$ as its generating
set. This means that $G_1$ has the elements of $G_a$ as vertices
and there is an edge from the vertex representing $g\in G_a$ to
the vertex representing $h\in G_a$ if for some $s\in S_a$, $g*s =
h$, where `$*$' denotes the group operation. If the generating
set $S_a$ is symmetric, i.e., if $a\in S_a$ implies $a^{-1}\in
S_a$, then the Cayley graph is undirected.

Let the two components of our (zig-zag product) graph be Cayley
graphs of the type $G_1=C(G_a,S_a)$ and $G_2=C(G_b,S_b)$ and
further, let us assume that there is a well-defined group action
by the group $G_b$ on the elements of the group $G_a$.  For $g
\in G_a$ and $h \in G_b$, let $g^{h}$ denote the action of $h$ on
$g$. Then the product graph is again a Cayley graph. More
specifically, if $G_1=C(G_a,S_a)$ and $G_2=C(G_b,S_b)$, and if
$S_a$ is the orbit of $k$ elements $a_1, a_2,\dots, a_k \in G_a$
under the action of $G_b$, then the generating set $S$ for the
Cayley (zig-zag product) graph is
$$
S=\{(1_{G_a},\beta)(a_i,1_{G_b})(1_{G_a},\beta')|\
\beta,\beta' \in S_b, i\in 1,\dots,k\}. $$

The group having as elements the ordered pairs $\{ (g,h)| g\in
G_a, h \in G_b \}$, and group operation defined by \[
(g',h')(g,h) = (g'g^{h'^-1} ,h'h)\] is called the {\em
  semi-direct product} of $G_a$ and $G_b$, and is denoted by
$G_a\rtimes G_b$. It is easily verified that when $k=1$, the
Cayley graph $C(G_a\rtimes G_b, S)$ is the zig-zag product
originally defined in~\cite{re02}. The degree of this Cayley
graph is at most $k|S_b|^2$ if we disallow multiple edges between
vertices. When the group sizes $G_a$ and $G_b$ are large and the
$k$ distinct elements $a_1, a_2,\dots, a_k \in G_a$ are chosen
randomly, then the degree of the product graph is almost always
$k|S_b|^2$.

\vspace{0.1in}

\subsection{Replacement product}
Let $G_1$ be a $(N_1,d_1,\lambda^{(1)})$-graph and let $G_2$ be a
$(d_1,d_2,\lambda^{(2)})$-graph.
Randomly number the edges around each vertex of $G_1$ by
$\{1,\ldots,d_1\}$, and each vertex of $G_2$ by $\{1,\ldots,d_1\}$.
Then the replacement product $G = G_1 \circr G_2$ of $G_1$ and $G_2$
has vertex set and edge set defined as follows: the vertices of $G$
are represented as ordered two tuples $(v,k)$, for $v
\in\{1,2,\dots,N_1\}$ and $k\in\{1,2,\dots,d_1\}$. There is an edge
between $(v,k)$ and $(v,\ell)$ if there is an edge between $k$ and
$\ell$ in $G_2$; there is also an edge between $(v,k)$ and
$(w,\ell)$ if the $k^{th}$ edge incident on vertex $v$ in $G_1$ is
connected to vertex $w$ and this edge is the $\ell^{th}$ edge
incident on $w$ in $G_1$. Note that the degree of the replacement
product graph depends only on the degree of the smaller component
graph $G_2$. The replacement product graph $G=G_1\circr G_2$ is a
$(N_1\cdot d_1, d_2+1,\lambda)$-graph with $\lambda \le
(p+(1-p)f(\lambda^{(1)},\lambda^{(2)}))^{1/3}$ for
$p=d_2^2/(d_2+1)^3$, where
$f(\lambda^{(1)},\lambda^{(2)})=\lambda^{(1)}+\lambda^{(2)}+[\lambda^{(2)}]^2$~\cite[Theorem
6.4]{re02}.

\begin{figure}[ht]
\centering{\resizebox{5in}{4in}{\includegraphics{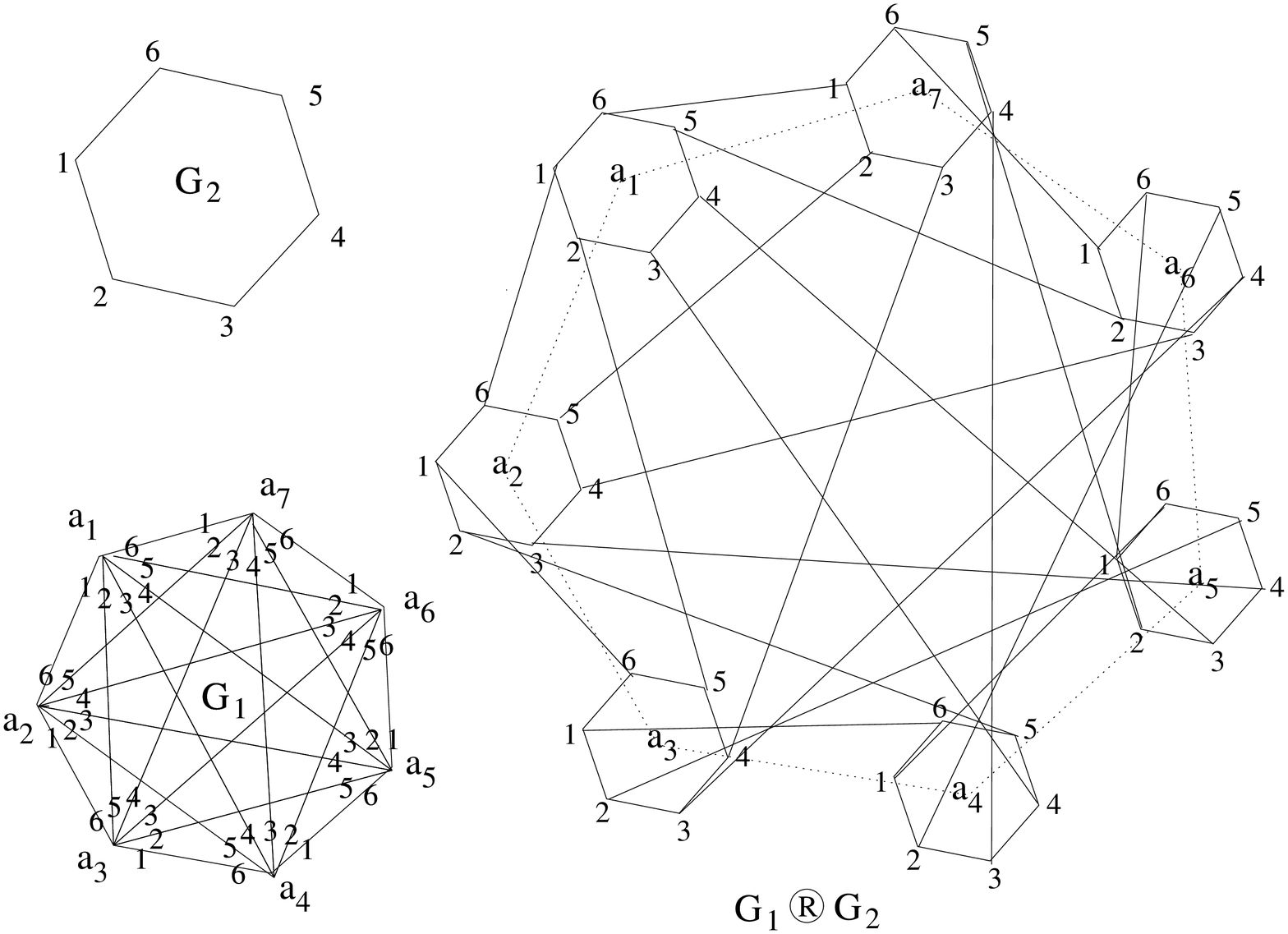}}}
\caption{Replacement product of two graphs.}
\label{replacement_prod}
\end{figure}
\vspace{0.1in}

\begin{example}
  Consider the replacement product graph $G = G_1 \circr G_2$
  shown in Figure 3. The vertex $(1,6)$ (``cloud $a_1$, vertex
  6") has degree $d_2+1$ = 3. The edges from $(1,6)$ to $(1,1)$
  and $(1,5)$ result since in $G_2$, vertex 6 connects to
  vertices 1 and 5. The edge from $(1,6)$ to $(7,1)$ result since
  in $G_1$, $a_1$ is the first neighbor of $a_7$ and $a_7$ is the
  sixth neighbor of $a_1$ in the labeling of $G_1$. Similarly,
  $(1,5)$ connects to $(1,6)$ and $(1,4)$ due to the original
  connections in $G_2$, and $(1,5)$ has an edge to $(6,2)$ since
  $a_1$ is the second neighbor of $a_6$ and $a_6$ is the fifth
  neighbor of $a_2$.
\end{example}
\vspace{0.1in}

\begin{lemma} Let $G_1$ (resp., $G_2$) have girth $g_1$ and diameter $t_1$(resp.,
  $g_2$, $t_2$). Then the girth $g$ and diameter $t$ of the replacement product graph
$G=G_1\circr G_2$ are given by: (a) girth $\min\{g_2,2g_1 \}\le g
\le \min\{g_2,g_1t_2\}$, and (b) diameter $ \max\{t_2, 2t_1\} \le
t\le t_1+t_2$. \label{rp_lemma}
\end{lemma}
\vspace{0.1in}

\begin{proof}
(a) Observe that there are cycles of length $g_2$ in $G$, as $G_2$
is a subgraph of $G$. Moreover, consider two vertices in $G_1$ on a
cycle of length $g_1$. Their clouds are $g_1$ apart in $G$, so a
smallest cycle between them would contain at most $g_1t_2$ edges (in
the worst case, $t_2$ steps would be needed within each cloud in the
$G_1$-cycle). So $g \le \min\{g_2, g_1t_2\}$. For the
  lower bound, the smallest cycle possible involving
  vertices in different clouds has length $2g_1$, and would occur
  if in the cycle, only one step was needed on each cloud. Thus,
  $g \ge \min\{g_2, 2g_1\}$.
(b) For the diameter, the furthest two vertices could be would occur
if they belonged to clouds associated to vertices at distance $t_1$
apart in $G_1$, and the path between them in $G$ would require at
most $t_2$ steps on each cloud. Therefore, $t \le t_1t_2$.
Similarly, the furthest distance between vertices in the same cloud
is $t_2$, and the furthest distance between vertices in different
clouds is at least $2t_1$, which would occur if they lie in clouds
associated to vertices at distance $t_1$ apart in $G_1$, but only
one step was needed on each cloud on the path. So $t \ge \max\{t_2,
2t_1\}$.
\end{proof}
\vspace{0.1in}

As earlier, let the two components of the product graph be Cayley
graphs of the type $G_1=C(G_a,S_a)$ and $G_2=C(G_b,S_b)$ and again
assume that there is a well-defined group action by the group $G_b$
on the elements of the group $G_a$. Then the replacement product
graph is again a Cayley graph. If $S_a$ is the union of $k$ orbits,
i.e., the orbits of $a_1, a_2, \dots, a_k \in G_a$ under the action
of $G_b$, then the replacement product graph is the Cayley graph of
the semi-direct product group $G_a\rtimes G_b$ and has
$S=(1_{G_a},S_b) \bigcup \{(a_1,1_{G_b}),\dots,(a_k,1_{G_b})\}$ as
the generating set.
The degree of this Cayley graph is $|S_b|+k$ and the size of its
vertex set is $|G_a||G_b|$~\cite{li03r}. (Here again, it is easily
verified that when $k=1$, the Cayley graph $C(G_a\rtimes G_b, S)$ is
the replacement product originally defined in~\cite{li03r}.)

\section{Zig-zag product for unbalanced bipartite graphs}
                               \label{sec-4}

For the purpose of coding theory it would be very interesting to
have a product construction of good unbalanced bipartite expanders.
In this Section we adapt the original zig-zag construction in a
natural manner. The main result (Theorem~\ref{zztheorem}) will show
that this construction results in a bipartite expander graph if the
component bipartite graphs are expander graphs.

Let $G_1$ be a $(c_1,d_1)$-regular graph on the vertex sets $V_1,
W_1$, where $|V_1| = N$ and $|W_1| = M$.  Let $G_2$ be a
$(c_2,d_2)$-regular graph on the vertex sets $V_2, W_2$, where
$|V_2| = d_1$ and $|W_2| = c_1$. Let $\lambda^{(1)}$ and
$\lambda^{(2)}$ denote the second largest eigenvalues of the
normalized adjacency matrices of $G_1$ and $G_2$, respectively.
Again, randomly number the edges around each vertex $\tilde{v} $
in $G_1$ and $G_2$ by $\{1,\ldots, deg(\tilde{v})\}$, where
$deg(\tilde{v})$ is the degree of $\tilde{v}$. Then the zig-zag
product graph, which we will denote by $G=G_1\circzb G_{2}$, is a
$(c_2^2,d_2^2)$-regular bipartite graph on the vertex sets $V, W$
with $|V|=N\cdot d_1$, $|W|=M\cdot c_1$, formed in the following
manner:

\begin{figure}[h]
  \centering{\resizebox{5in}{3in}{\includegraphics{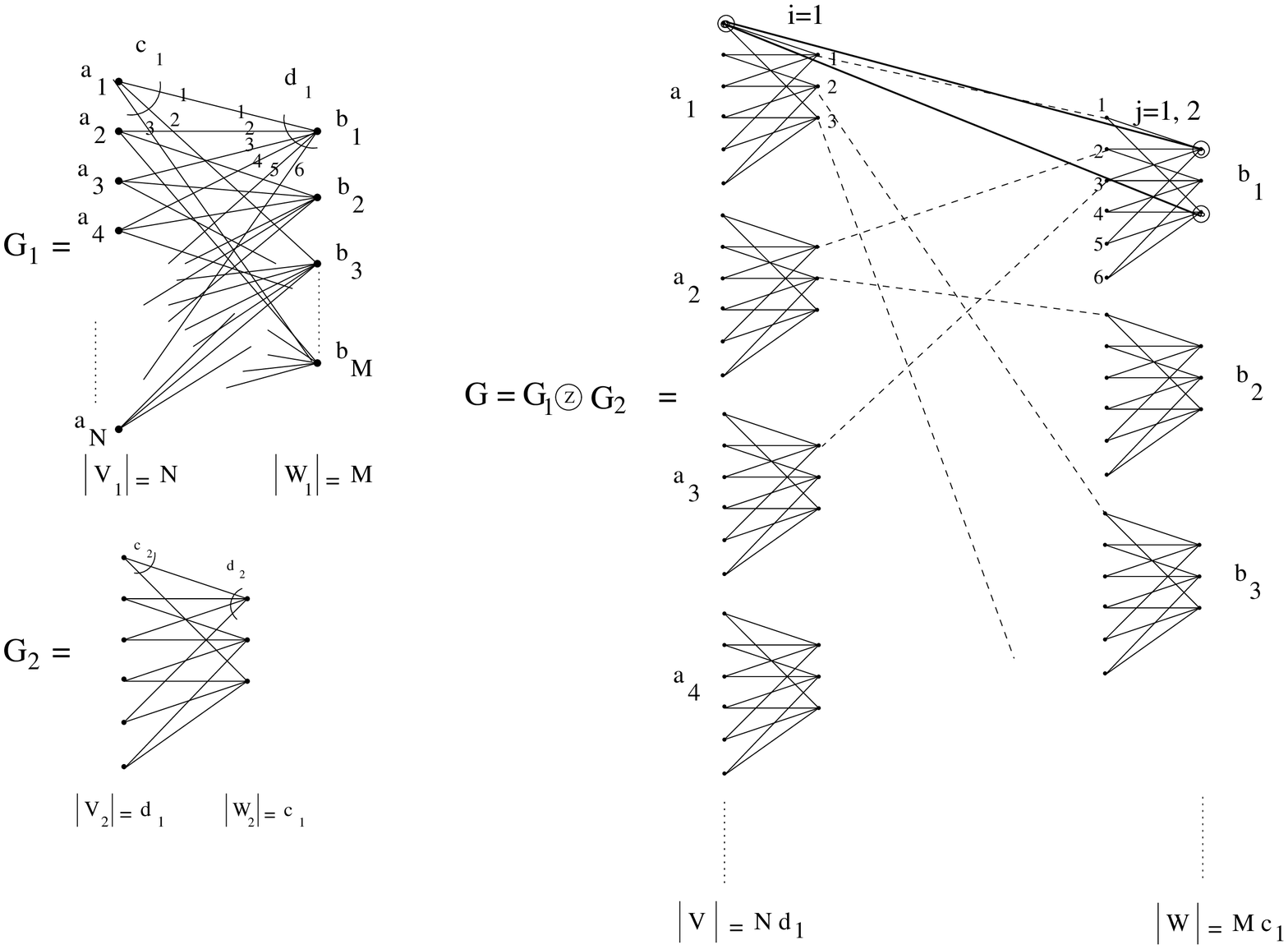}}}
\caption{Zig-Zag product of two unbalanced bipartite graphs.}
\label{zigzag_bipartiteU1}
\end{figure}

\vspace{0.1in}
\begin{itemize}
\item Every vertex $v \in V_1$ and $w\in W_1$ of $G_1$ is
  replaced by a copy of $G_2$. The cloud at a vertex $v \in V_1$
  has vertices $V_2$ on the left and vertices $W_2$ on the right,
  with each vertex from $W_2$ corresponding to an edge from $v$
  in $G_1$. The cloud at a vertex $w\in W_1$ is similarly
  structured with each vertex in $V_2$ in the cloud corresponding
  to an edge of $w$ in $G_1$. (See
  Figure~\ref{zigzag_bipartiteU1}.)  Then the vertices from $V$
  are represented as ordered pairs $(v,k)$, for $v\in
  \{1,\dots,N\}$ and $k\in \{1,\dots,d_1\}$, and the vertices
  from $W$ are represented as ordered pairs $(w,\ell)$, for $w\in
  \{1,\dots,M\}$ and $\ell\in \{1,\dots,c_1\}$.

\item A vertex $(v,k) \in V$ is connected to a vertex in $W$ by
  making three steps in the product graph:

\begin{itemize}
\item A small step ``zig'' from left to right in the local copy
  of $G_2$. This is a step $(v,k) \rightarrow (v, k[i])$, for
  $i\in\{1,\dots,c_2\}$.
\item A deterministic step from left to right on $G_{1}$
  $(v,k[i]) \rightarrow (v[k[i]],\ell)$, where $v[k[i]]$ is the
  $k[i]^{th}$ neighbor of $v$ in $G_1$ and $v$ is the $\ell^{th}$
  neighbor of $v[k[i]]$ in $G_1$.
\item A small step ``zag'' from left to right in the local copy
  of $G_{2}$. This is a step $(v[k[i]],\ell) \rightarrow
  (v[k[i]],\ell[j])$, where the final vertex is in $W$, for $j\in
  \{1,\dots,c_2\}$.
\end{itemize}

Therefore, there is an edge between $(v,k)$ and
$(v[k[i]],\ell[j])$.
\end{itemize}
\vspace{0.1in}

There is a subtle difference in the zig-zag product construction
described for the unbalanced bipartite component graphs when
compared to the original construction in~\cite{re02}. The
difference lies in that the vertex set of $G$ does not include
vertices from the set $W_2$ in every cloud of vertices from $V_1$
and similarly, the vertex set of $G$ does not include vertices
from $V_2$ in every cloud of vertices from $W_1$.

The following theorem describes the major properties of the
constructed unbalanced bipartite graph.

\begin{theorem}                              \label{zztheorem}
  Let $G_1$ be a $(c_1,d_1)$-regular bipartite graph on $(N, M)$
  vertices with $\lambda(G_1)=\lambda^{(1)}$, and let $G_2$ be a
  $(c_2,d_2)$-regular bipartite graph on $(d_1, c_1)$ vertices
  with $\lambda(G_2)=\lambda^{(2)}$. Then, the zig-zag product
  graph $G_1\circzb G_2$ is a $(c_2^2,d_2^2)$-regular bipartite
  on $(N\cdot d_1, M\cdot c_1)$ vertices with
  $\lambda=\lambda(G_1\circzb G_2) \le
  \lambda^{(1)}+\lambda^{(2)}+[\lambda^{(2)}]^2$. Moreover, if
  $\lambda^{(1)}<1$ and $\lambda^{(2)}<1$, then
  $\lambda=\lambda(G_1\circzb G_2) < 1$.
\end{theorem}
\vspace{0.1in}

The proof of the expansion of the unbalanced zig-zag product
graph is nontrivial and will require the remainder of
Section~\ref{sec-4}. Several of the key ideas in the
following proof are already present in the original zig-zag
product graph paper~\cite{re02}. Some modifications for balanced
bipartite graphs has been dealt in~\cite{li03r}. Note that unlike
in the original zig-zag product construction~\cite{re02}, the
vertex set of $G$ does not include vertices from the set $W_2$ in
any cloud of vertices from $V_1$, nor vertices from $V_2$ in any
cloud of vertices from $W_1$.  However, the girth of the
unbalanced zig-zag product is also 4, and this can be seen using
a similar argument as in Lemma~\ref{zz_lemma}.

\begin{proof}
  Let $M_G$ denote the adjacency matrix of $G$. For convenience,
  we also let $G_2$ denote the $d_1\times c_1$ matrix that
  describes the connections between the nodes in $V_2$ to the
  nodes in $W_2$ for the graph $G_2$, and $G_1$ denote the
  $N\times M$ matrix that describes the connections between the
  nodes in $V_1$ and the nodes in $W_1$ for the graph $G_1$. This
  means that the adjacency matrix for the graph $G_2$ is given by
  $M_2 = \left[
    \begin{array}{cc}
      0&G_2\\
      G_2^T&0
\end{array}
\right]$, and the adjacency matrix for the graph $G_1$ is given
by $M_1 = \left[ \begin{array}{cc}
    0&G_1\\
    G_1^T&0
\end{array}
\right]$.

The adjacency matrix for the zig-zag product graph $G$ is given
by
\[M_G =  \left[ \begin{array}{cc}
0& (G_2\otimes I_n){\tilde{A}_2}(G_2^T\otimes I_m)\\
(G_2^T\otimes I_m){\tilde{A}_2}(G_2\otimes I_n)&0
\end{array} \right], \] where $\tilde{A}_2$ is a permutation matrix
of size $Nc_1\times Nc_1$ that describes the zig-zag product
connections.

The largest eigenvalue of $M_G$ is $c_2 d_2$ and the
corresponding eigenvector is $v_0=[1 \ \cdots \ 1 \ r\ \cdots \
r]^T$, where the first $Nd_1$ components are equal to 1 and the
remaining $Mc_1$ components are equal to $r =\frac{d_2}{c_2}
=\frac{d_1}{c_1}$.

Let $1_x$ denote a column vector of length $x$ with all entries
equal to 1. Then, the largest eigenvalue of $M_1$ is
$\sqrt{c_1d_1}$ and the corresponding eigenvector is $w_0= \left[
\begin{array}{c} 1_{N}\\
r_11_{M}\\
\end{array}
\right]$, where $r_1=\sqrt{\frac{d_1}{c_1}}=\sqrt{r}$.
Similarly, the largest eigenvalue of $M_2$ is $\sqrt{c_2d_2}$ and
the corresponding eigenvector is $u_0 = \left[
\begin{array}{c} 1_{d_1}\\
r_21_{c_1}\\
\end{array}
\right]$, where $r_2=\sqrt{\frac{d_2}{c_2}}=r_1=\sqrt{r}$.

Let $\lambda_2 = \max_{u \perp u_0} \frac{<M_2u,u>}{<u,u>} =
\max_{u \perp u_0} { \frac{2(u_a^TG_2u_b)}{\parallel u_a
    \parallel^2 + \parallel u_b \parallel^2}}$, where $u = \left[
  \begin{array}{c}
    u_a\\
    u_b
\end{array}
\right]$.  That is $u_a$ is a column vector of length $d_1$
corresponding to the vertices in $V_2$ and $u_b$ is a column
vector of length $c_1$ corresponding to the vertices in $W_2$.
We choose $u$ such that $u \perp u_0$.  Furthermore, $u$ can be
written as two vectors $u^{\parallel}$ and $u^{\perp}$ where
$u^{\parallel}$ is a vector that is parallel to the constant
(non-zero) vector and $u^{\perp}$ is a vector that is
perpendicular (or, orthogonal) to the constant (non-zero) vector.
That is $u = u^{\parallel}+u^{\perp}$.

(Note that, by definition, $\lambda_2$ corresponds to the second
largest eigenvalue of $M_2$ and the corresponding eigenvector $u$
that maximizes $\lambda_2$ in the above is orthogonal to the
$u_0$, the eigenvector corresponding to the eigenvalue
$\sqrt{c_2d_2}$.  )

Similarly, let $\lambda_1 = \max_{w \perp w_0}
\frac{<M_1w,w>}{<w,w>} = \max_{w\perp w_0} {
  \frac{2(w_a^TG_1w_b)}{\parallel w_a \parallel^2 + \parallel w_b
    \parallel^2}}$, where $w = \left[ \begin{array}{c}
    w_a\\
    w_b
\end{array}
\right]$. (Here, $w_a$ is a column vector of length $N$ and $w_b$
is a column vector of length $M$. $w$ can also be broken down as
$w=w^{\parallel}+w^{\perp}$ as above. By definition, $\lambda_1$
is the second largest eigenvalue of $M_1$.)

The eigenvector of $M_G$ corresponding to the largest eigenvalue
$c_2d_2$ is $v_0 = \left[ \begin{array}{c}
    1_{Nd_1}\\
    r1_{Mc_1}
\end{array}
\right]$. Let $\alpha = \left[ \begin{array}{c}
    \alpha_a\\
    \alpha_b
\end{array}
\right]$ be an eigenvector of $M_G$, where $\alpha_a$ has length
$Nd_1$ and $\alpha_2$ has length $Mc_1$. Let $e_m$ be a basis
vector with component value $1$ at the $m$th entry and component
value $0$ elsewhere. Then, the vectors $\alpha_a$ and $\alpha_b$
can be written as $\alpha_a = \sum_{n \in [N]} \alpha_{a_n}
\otimes e_n$, and $\alpha_b = \sum_{m \in [M]} \alpha_{b_m}
\otimes e_m$, where $\otimes$ denotes the Kronecker product,
$\alpha_{a_n}$ is the vector $\alpha_a$ restricted to the
components corresponding to the vertices in the $n^{th}$ vertex
cloud, $n\in [N]$, of the graph $G$, and $\alpha_{b_m}$, for
$m\in [M]$, is defined similarly.

Then the second largest eigenvalue of $M_G$ is $\lambda =
\max_{\alpha \perp v_0}
\frac{2\alpha_a^T(K)\alpha_b}{<\alpha,\alpha>}$, where $K = (G_2
\otimes I_N)\tilde{A}_2(G_2 \otimes I_M)$.

Let $s = |2(\alpha_a^TK\alpha_b)|$. Splitting $\alpha_{a_n}$
(and, $\alpha_{b_m}$) into parallel and perpendicular parts,
$\alpha_{a_n}=\alpha_{a_n}^{\parallel}+\alpha_{a_n}^{\perp}$, we
can write
\begin{multline*}
  s = 2\Big{(}\sum_{n \in [N]}((\alpha_{a_n}^{\parallel})^T
  G_2 \otimes e_n)\\
  + \sum_{n \in [N]}(\alpha_{a_n}^{\perp})^T G_2 \otimes
  e_m\Big{)} \tilde{A}_2 \Big{(}\sum_{m\in [M]} G_2
  \alpha_{b_m}^{\parallel} \otimes e_m + \sum_{m \in [M]} g_2
  \alpha_{b_m}^{\perp} \otimes e_m\Big{)}
\end{multline*}

We want to show that $s \le f(\lambda_1, \lambda_2)(\parallel
\alpha_a \parallel^2 + \parallel \alpha_b \parallel^2)$, where
$f(\lambda_1,\lambda_2)$ is some positive-valued function such
that $f(\lambda_1,\lambda_2)\le \lambda_1+\lambda_2+\lambda_2^2$.
Note that $\alpha_a \in \mathbb{R}^{Nd_1}$, and $\alpha_b \in
\mathbb{R}^{Mc_1}$.

Observe the following:
\begin{enumerate}
\item $\lambda_2 = \max_{u \perp {\scriptscriptstyle\left[
        \begin{array}{c}
          1_{d_1}\\
          \sqrt{r}1_{c_1}
\end{array}
\right]}} \frac{2 u_a^T g_2 u_b}{\parallel u_a \parallel^2 +
\parallel u_b \parallel^2},$ where $u = \left[ \begin{array}{c}
  u_a\\
  u_b
\end{array}
\right]$.

\item $\lambda_1 = \max_{u \perp {\scriptstyle \left[
        \begin{array}{c}
          1_{N}\\
          \sqrt{r}1_{M}
\end{array}
\right]}} \frac{2 w_a^T g_1 w_b}{ \parallel w_a \parallel^2 +
\parallel w_b \parallel^2},$ where $w = \left[ \begin{array}{c}
  w_a\\
  w_b
\end{array}
\right]$.

\item $\alpha_a = \left[ \begin{array}{c}
      \alpha_{a_1}\\
\vdots\\
\alpha_{a_N}
\end{array}
\right]$, where $ \alpha_{a_i} \in \mathbb{R}^{d_1}, i \in [N]$,
$\alpha_b = \left[
\begin{array}{c}
\alpha_{b_1}\\
\vdots\\
\alpha_{b_M}
\end{array}
\right]$, \\
where $ \alpha_{b_i} \in \mathbb{R}^{c_1}, i \in [M]$, and
$\alpha = \left[ \begin{array}{c}
    \alpha_a\\
    \alpha_b
\end{array}
\right] \in \mathbb{R}^{Nd_1+Mc_1}$.

\item From the definition of $\lambda_2$, we have
\[ \parallel \left[ \begin{array}{cc}
0&G_2\\
G_2^T&0
\end{array}
\right] \left[ \begin{array}{c}
    \alpha_{a_n}\\
    0
\end{array}
\right] \parallel \le \lambda_2 |\left[ \begin{array}{c}
    \alpha_{a_n}\\
    0
\end{array}
\right]| \]

\[ \Rightarrow \parallel G_2^T\alpha_{a_n} \parallel \le \lambda_2 \parallel
\alpha_{a_n}^{\perp} \parallel \mbox{ and } \parallel
G_2^T\alpha_{a_n} \parallel \le \lambda_2 \parallel \alpha_{a_n}
\parallel.\]

This implies that $\left[ \begin{array}{c}
    \alpha_{a_n}\\
    0
\end{array}
\right] \perp \left[ \begin{array}{c}
    1_{d_1}\\
    \sqrt{r}1_{c_1}
\end{array}
\right]$.

\item Since $\alpha_{a_n}^{\perp}$ is orthogonal to the constant
  vector, we have $\alpha_{a_n}^{\perp} \perp 1_{d_1}$.
\end{enumerate}
\vspace{0.3in}

Rewriting, we have that
\begin{multline*}
  s = 2\Big{(}\sum_{n \in [N]} (\alpha_{a_n}^{\parallel})^TG_2
  \otimes e_n\Big{)}\tilde{A}_2 \Big{(}\sum_{m
    \in [M]} G_2\alpha_{b_m}^{\parallel} \otimes e_m\Big{)}\\
  + 2\Big{(}\sum_{n \in [N]}(\alpha_{a_n}^{\parallel})^TG_2
  \otimes e_n\Big{)}\tilde{A}_2\Big{(}\sum_{m \in
    [M]} G_2\alpha_{b_m}^{\perp} \otimes e_m\Big{)}\\
  + 2\Big{(}\sum_{n \in [N]}(\alpha_{a_n}^{\perp})^TG_2 \otimes
  e_n\Big{)}\tilde{A}_2 \Big{(}\sum_{m \in
    [M]} G_2\alpha_{b_m}^{\parallel} \otimes e_m\Big{)}\\
  +2\Big{(}\sum_{n \in [N]}(\alpha_{a_n}^{\perp})^TG_2 \otimes
  e_n\Big{)}\tilde{A}_2 \Big{(}\sum_{m \in [M]}
  G_2\alpha_{b_m}^{\perp} \otimes e_m\Big{)}.
\end{multline*}

So $s = s_1+ s_2+s_3+s_4$, where
\[ s_1  = 2\Big{(}\sum_{n \in [N]} (\alpha_{a_n}^{\parallel})^TG_2 \otimes
e_n\Big{)}\tilde{A}_2 \Big{(}\sum_{m \in [M]}
G_2\alpha_{b_m}^{\parallel} \otimes e_m\Big{)}\]
\[ s_2 = 2\Big{(}\sum_{n \in [N]}(\alpha_{a_n}^{\parallel})^TG_2 \otimes
e_n\Big{)}\tilde{A}_2\Big{(}\sum_{m \in [M]}
G_2\alpha_{b_m}^{\perp} \otimes e_m\Big{)}\]
\[ s_3 = 2\Big{(}\sum_{n \in [N]}(\alpha_{a_n}^{\perp})^TG_2 \otimes
e_n\Big{)}\tilde{A}_2 \Big{(}\sum_{m \in [M]}
G_2\alpha_{b_m}^{\parallel} \otimes e_m\Big{)}, \mbox{ and }\]
\[ s_4 = 2\Big{(}\sum_{n \in [N]}(\alpha_{a_n}^{\perp})^TG_2 \otimes
e_n\Big{)}\tilde{A}_2 \Big{(}\sum_{m \in [M]}
G_2\alpha_{b_m}^{\perp} \otimes e_m\Big{)}. \]

\vspace{0.2in}

We will bound each part of $s$ separately:

\begin{enumerate}

\item Since $\tilde{A}_2$ is a permutation matrix, we have \[s_4
  \le 2 \parallel \sum_{n\in[N]}(\alpha_{a_n}^{\perp})^TG_2
  \otimes e_n \parallel \parallel \sum_{m\in[M]}G_2
  \alpha_{b_m}^{\perp} \otimes e_m \parallel\] Since $\parallel
  (\alpha_{a_n}^{\perp})^TG_2\parallel \le \lambda_2 \parallel
  \alpha_{a_n}^{\perp}\parallel $ by definition of $\lambda_2$
  (and similarly,\\ $\parallel G_2\alpha_{b_m}^{\perp}\parallel
  \le \lambda_2 \parallel \alpha_{b_n}^{\perp}\parallel $), we
  have \[s_4 \le 2 \parallel \sum_{n\in [N]} \lambda_2
  \alpha_{a_n}^{\perp} \otimes e_n \parallel \parallel \lambda_2
  \sum_{m\in [M]} \alpha_{b_m}^{\perp} \otimes e_m \parallel \]
\[
= 2\lambda_2^2 \parallel \alpha_a^{\perp} \parallel \parallel
\alpha_b^{\perp}\parallel \le \lambda_2^2(\parallel
\alpha_a^{\perp} \parallel^2+ \parallel
\alpha_b^{\perp}\parallel^2)=\lambda_2^2(\parallel
\alpha^{\perp}\parallel^2)
\]

\item Since $\tilde{A}_2$ is a permutation matrix, we have
\[
s_3 \le 2 \parallel \sum_{n\in[N]} (\alpha_{a_n}^{\perp})^TG_2
\otimes e_n\parallel \parallel \sum_{m\in[M]} G_2
\alpha_{b_m}^{\parallel} \otimes e_m\parallel
\]
Using the argument from the previous step and since $\parallel
G_2 \alpha_{b_m}^{\parallel}\parallel \le \parallel
\alpha_{b_m}^{\parallel}\parallel $, we have
\[s_3 \le
2\lambda_2 \parallel \alpha_a^{\perp}\parallel \parallel
\alpha_b^{^{\parallel}}\parallel\]

\item Similarly, we can show that
 \[ s_2 \le 2\lambda_2 \parallel \alpha_a^{\parallel}\parallel
 \parallel\alpha_b^{\perp}\parallel \]

\item To upper bound $s_1$, define a new vector $C(\alpha_a')$
  for every vector $\alpha_a'$ such that its $m$th component is
 \[(C(\alpha_a'))_m := \frac{1}{d_1}\sum_{a=1}^{d_1} \alpha_{a'_m}, \mbox{
   for } m\in[M] \] This implies that $ \alpha_a'^{\parallel} =
 C(\alpha_a') \otimes \frac{1d_1}{d_1}.$

 Similarly, define a new vector $C'(\alpha_b')$ for every
 $\alpha_b'$ as \[(C'(\alpha_b'))_n :=
 \frac{1}{c_1}\sum_{b=1}^{c_1} \alpha_{b_n}', \mbox{ for }
 n\in[N] \] This implies that $ \alpha_{b}'^{\parallel} =
 C'(\alpha_b') \otimes \frac{1c_1}{c_1}.$

 That is, the functions $C(\cdot)$ and $C'(\cdot)$ computes the
 average value of the components in each vertex cloud of the
 zig-zag product graph $G$.

 Therefore, we have $C'\tilde{A}_2(e_m \otimes \frac{1d_1}{d_1})
 = G_1 e_m $.  Note that $\alpha_{b_m}^{\parallel}G_2 =
 \alpha_{\tilde{a}_m}^{\parallel}$, where $\tilde{a}_m$ in the
 subscript refers to the left vertices on the right of the
 zig-zag product graph that are used for the construction but do
 not belong to the vertex set of the zig-zag product graph.
 Rewriting $s_1$, we have


\begin{multline*}
  \hspace*{5mm} s_1 = 2(\sum_{n\in[N]}
  (\alpha_{a_n}^{\parallel})^T G_2 \otimes
  e_n)^T\tilde{A}_2(\sum_{m\in[M]}
  \alpha_{b_m}^{\parallel} G_2\otimes e_m) \\
  = 2(\sum_{n\in[N]} \alpha_{\tilde{b}_n}^{\parallel} \otimes
  e_n)^T \tilde{A}_2 (\sum_{m \in [M]}
  \alpha_{\tilde{a}_m}^{\parallel} \otimes e_m).
\end{multline*}


This is because, $(\alpha_{a_n}^{\parallel})^T
G_2=\alpha_{\tilde{b}_n}^{\parallel}$, the components of the
right vertices of the $G_2$ clouds on the left of $G$, and
$G_2(\alpha_{b_m}^{\parallel})=\alpha_{\tilde{a}_m}^{\parallel}$,
the components corresponding to the left vertices of the $G_2$
clouds on the right of $G$.  (That is, since $G_2$ denotes the
connections between the left vertices and right vertices,
multiplying with $G_2$ takes $(\alpha_{a_n}^{\parallel})^T$ to
$(\alpha_{\tilde{b}_n}^{\parallel})$ and
$(\alpha_{b_m}^{\parallel})$ to
$(\alpha_{\tilde{a}_m}^{\parallel})$.)

But $\alpha_a^{\parallel} = (C(\alpha_a)) \otimes
\frac{1d_1}{d_1}$, $\alpha_{\tilde{b}}^{\parallel} =
(C'(\alpha_{\tilde{b}}) \otimes \frac{1c_1}{c_1}$. Hence,
\[
s_1 = 2(C'(\alpha_{\tilde{b}}) \otimes
\frac{1c_1}{c_1})^T\tilde{A}_2(C(\alpha_{\tilde{a}})\otimes\frac{1d_1}{d_1})\]
\[\Rightarrow s_1 = 2(C(\alpha_{\tilde{b}}))^TG_1(C(\alpha_{\tilde{a}}))/(c_1 d_1)
=\frac{2(\frac{1}{\sqrt{c_1}}C'(\alpha_{\tilde{b}}))^TG_1
  (\frac{1}{\sqrt{d_1}}C(\alpha_{\tilde{a}}))}{\sqrt{c_1}\sqrt{d_1}}
\]

But observe that $\left[ \begin{array}{c}
    \frac{1}{\sqrt{c_1}}C'(\alpha_{\tilde{b}})\\
    \frac{1}{\sqrt{d_1}}C(\alpha_{\tilde{a}})
\end{array}
\right]$ is orthogonal to the vector $\left[ \begin{array}{c}
    1_N\\
    \sqrt{r}1_M
\end{array}
\right]$.  This is because \[ <\left[ \begin{array}{c}
    \frac{1}{\sqrt{c_1}}C'(\alpha_{\tilde{b}})\\
    \frac{1}{\sqrt{d_1}}C(\alpha_{\tilde{a}})
\end{array}\right]
,\left[ \begin{array}{c}
    1_N\\
    \sqrt{r}1_M
\end{array}\right] > =
\frac{1}{\sqrt{c_1}}\sum_{b=1}^{c_1}\sum_{n=1}^N
\alpha_{{\tilde{b}}_n}
+\frac{1}{\sqrt{d_1}}\frac{\sqrt{d_1}}{\sqrt{c_1}}\sum_{a=1}^{d_1}\sum_{m=1}^M
\alpha_{{\tilde{a}}_m}\ \ (**)\] However,
$\sum_{b=1}^{c_1}\sum_{n=1}^N
\alpha_{b_n}=c_2\sum_{a=1}^{d_1}\sum_{n=1}^N \alpha_{a_n}$
and \\
$\sum_{a=1}^{d_1}\sum_{m=1}^M
\alpha_{a_m}=d_2\sum_{b=1}^{c_1}\sum_{m=1}^M \alpha_{b_m}$.
Since $\left[ \begin{array}{c}
    \alpha_a\\
    \alpha_b
\end{array}\right]$ was chosen to be orthogonal to $\left[ \begin{array}{c}
1_{Nd_1}\\
r 1_{Mc_1}
\end{array}\right]$, it is easy to verify that the sum in $(**)$ is zero.

Hence, from the definition of $\lambda_1$, we have
\[
s_1=\frac{2(\frac{1}{\sqrt{c_1}}C'(\alpha_{\tilde{b}}))^TG_1(\frac{1}{\sqrt{d_1}}C
  (\alpha_{\tilde{a}}))}{{c_1}{d_1}} \le
\frac{\lambda_1}{{c_1d_1}}(\frac{1}{c_1}\parallel
C'(\alpha_{\tilde{b}})\parallel^2 + \frac{1}{d_1}\parallel
C(\alpha_{\tilde{a}})\parallel^2) \ \ \ (*)
\]

It is easy to verify that the RHS in $(*)$ can be upper bounded
as
\[
s_1\le RHS (*) \le \lambda_1 (\parallel
\alpha_a^{\parallel}\parallel^2 +\parallel
\alpha_b^{\parallel}\parallel^2) =\lambda_1 (\parallel
\alpha^{\parallel}\parallel^2)
\]

%

%
%
%

\end{enumerate}

Combining the upper bounds on $s_1,s_2,s_3,s_4$, we have
\[s=s_1+s_2+s_3+s_4\]
\[s\le\lambda_1(\parallel
\alpha^{\parallel}\parallel^2)+2\lambda_2(\parallel
\alpha_a^{\perp}\parallel \parallel
\alpha_b^{\parallel}\parallel+\parallel
\alpha_a^{\parallel}\parallel \parallel \alpha_b^{\perp}\parallel
)+ \lambda_2^2(\parallel \alpha^{\perp}\parallel^2)\] However,
observe that
\[2\lambda_2(\parallel \alpha_a^{\perp}\parallel
\parallel \alpha_b^{\parallel}\parallel+\parallel
\alpha_a^{\parallel}\parallel \parallel
\alpha_b^{^{\perp}}\parallel ) \le \] \[ \lambda_2(\parallel
\alpha_a^{\parallel}\parallel^2+\parallel
\alpha_a^{\perp}\parallel^2+\parallel
\alpha_b^{\parallel}\parallel^2+\parallel
\alpha_b^{\perp}\parallel^2) =\lambda_2(\parallel \alpha
\parallel^2).\] Further, $\parallel \alpha^{\parallel}
\parallel^2 \le \parallel \alpha \parallel^2$ and $\parallel
\alpha^{\perp} \parallel^2 \le \parallel \alpha \parallel^2$.

Thus, we have
\[s\le (\lambda_1+\lambda_2+\lambda_2^2)(\parallel \alpha \parallel^2)\]

The second largest eigenvalue of $M_G$ is defined as $\lambda =
\max_{\alpha \perp v_0} \frac{s}{<\alpha,\alpha>}$, where $s =
2\alpha_a^T(G_2 \otimes I_N)\tilde{A}_2(G_2 \otimes
I_M)\alpha_b$.  Using the upper bound on $s$, we get $\lambda \le
\frac{(\lambda_1+\lambda_2+\lambda_2^2)(\parallel \alpha
  \parallel^2)}{\parallel \alpha \parallel^2} =
\lambda_1+\lambda_2+\lambda_2^2$.

The only remaining step is to show that if $\lambda_1 <1 $ and
$\lambda_2<1$, then $\lambda<1$.  Suppose $\lambda_1<1,
\lambda_2<1$ and suppose $\parallel \alpha^{\perp} \parallel \le
\frac{1-\lambda_1}{3\lambda_2}\parallel \alpha \parallel$.  Then,
we can upper bound $s$ as follows
\[
s\le \lambda_1 \parallel \alpha ^{\parallel}\parallel^2
+2\lambda_2 \parallel \alpha^{\parallel}\parallel \parallel
\alpha^{\perp}\parallel + \lambda_2^2\parallel
\alpha^{\perp}\parallel^2\]
\[\le \parallel \alpha \parallel^2 + \frac{2(1-\lambda_1)}{3}\parallel
\alpha \parallel^2 +\frac{(1-\lambda_1)^2}{9}\parallel \alpha
\parallel^2 = (1-\frac{1-\lambda_1}{3})^2 \parallel \alpha
\parallel^2 \le \parallel \alpha \parallel^2 \]

Suppose $\parallel \alpha^{\perp} \parallel >
\frac{1-\lambda_1}{3\lambda_2}\parallel \alpha \parallel$.

Then, notice that $s= 2(\alpha_a^{\parallel}+
\alpha_a^{\perp})(\sum_{n}G_2^T\otimes e_n) A_2 (\sum_m
G_2\otimes e_m) (\alpha_b^{\parallel}+\alpha_b^{\perp})$.  The
RHS can be written as
$2(\alpha_{\tilde{b}}^{\parallel}+\sum_n\alpha_{a_n}^{\perp}(G_2^T)\otimes
e_n)A_2(\alpha_{\tilde{a}}^{\parallel}+\sum_m
G_2\alpha_{b_m}^{\perp} \otimes e_m$.  However,
$\sum_n\alpha_{a_n}^{\perp}(G_2^T)\otimes e_n$ is orthogonal to
$\alpha_{\tilde{a}}^{\parallel}$ and $\sum_m G_2
\alpha_{b_m}^{\perp}\otimes e_m$ is orthogonal to
$\alpha_{\tilde{b}}^{\parallel}$.  Thus, \[s=
2(\alpha_{\tilde{b}}^{\parallel})A_2(\alpha_{\tilde{a}}^{\parallel})+
2(\sum_n\alpha_{a_n}^{\perp}(G_2^T)\otimes e_n )A_2 (\sum_m
G_2\alpha_{b_n}^{\perp}\otimes e_m )
\]
{}From the previous arguments, we have
\[ s \le \lambda_1(\parallel \alpha^{\parallel}\parallel^2)
+\lambda_2^2(\parallel \alpha^{\perp}\parallel^2)\]
\[= \lambda_1(\parallel \alpha \parallel^2-\parallel
\alpha^{\perp}\parallel^2) +\lambda_2^2\parallel \alpha^{\perp}
\parallel^2
\]
\[
\le (\parallel \alpha \parallel^2-\parallel
\alpha^{\perp}\parallel^2) +\lambda_2^2\parallel \alpha^{\perp}
\parallel^2 = (1-\frac{(1-\lambda_1)^2(1-\lambda_2^2)}{9})
\parallel \alpha \parallel^2 \le \parallel \alpha \parallel^2
\]

This completes the proof.
\end{proof}

\section{Zig-zag and replacement product LDPC codes}             \label{sec-5}
In this section, we design LDPC codes based on expander graphs
arising from the zig-zag and replacement products. The zig-zag
product of regular graphs yields a regular graph which may or may
not be bipartite, depending on the choice of the component
graphs.  Therefore, to translate the zig-zag product graph into a
LDPC code, the vertices of the zig-zag product are interpreted as
sub-code constraints of a suitable linear block code and the
edges are interpreted as code bits of the LDPC code. This is akin
to the procedure described in~\cite{ta81a} and~\cite{la00p}. The
same procedure is applied to the replacement product graphs.

We further restrict the choice of the component graphs for our
products to be appropriate Cayley graphs so that we can work
directly with the group structure of the Cayley graphs. The
following examples, the first two using Cayley graphs from {\rm
  \cite{al01p}, illustrate the code construction technique:
  \vspace{0.1in}

\begin{example}  Let
  $A=\mathbb{F}_2^p$ be the Galois field of $2^p$ elements for a
  prime $p$, where the elements of $A$ are represented as vectors
  of a $p$-dimensional vector space over $\mathbb{F}_2$. Let
  $B=\mathbb{Z}_p$ be the group of integers modulo $p$. (Further,
  let $p$ be chosen such that the element $2$ generates the
  multiplicative group $\mathbb{Z}_p^* = \mathbb{Z}_p-\{0\}$.)
  The group $B$ acts on an element ${\bf
    x}=(x_0,x_1,\dots,x_{p-1}) \in A$ by cyclically shifting its
  coordinates, i.e. $\phi_b({\bf
    x})=(x_b,x_{b+1},\dots,x_{b-1}),\ \forall b\in B$. Let us now
  choose $k$ elements $a_1,a_2,\dots,a_k$ randomly from $A$. The
  result in~\cite[Theorem 3.6]{al01p} says that for a random
  choice of elements $a_1, a_2,\dots, a_k$, the Cayley graph
  $C(A,\{a_1^B,a_2^B,\dots,a_k^B\})$ is an expander with high
  probability. (Here, $a_i^B$ is the orbit of $a_i$ under the
  action of $B$.) The Cayley graph for the group $B$ with the
  generators $\{\pm 1\}$ is the cyclic graph on $p$ vertices,
  $C(B,\{\pm 1\})$.

  \indent(a) The zig-zag product of the two Cayley graphs is the
  Cayley graph \[C(A\rtimes B, S =\{(0,\beta)(a_i,0)(0,\beta')| \
  \beta,\beta'=\pm 1, i=1,2,..,k\})\] on $N=2^p\cdot p$ vertices,
  where $A\rtimes B$ is the semi-direct product group and the
  group operation is $(a,b)(c,d)=(a+\phi_b(c), b+d)$, for $a,c\in
  A,\ b,d\in B$. This is a regular graph with
  degree\footnote{Depending on the choice of the $a_i$'s, the
    number of distinct elements in $S$ may be fewer than
    $k|S_B|^2$.} $d_g\le k|S_B|^2=4k$. If we interpret the
  vertices of the graph as sub-code constraints of a
  $[d_g,k_g,d_m]$ linear block code and the edges of the graph as
  code bits of the LDPC code, then the block length $N_{LD}$ of
  the LDPC code is $2^p\cdot p\cdot d_g/2$ and the rate of the
  LDPC code is
  \[
  r\ge \frac{N_{LD}-N(d_g-k_g)}{N_{LD}} =
  1-\frac{2(d_g-k_g)}{d_g}=\frac{2k_g}{d_g}-1.
  \]
  (Observe that $r\ge 2r_1-1$, where $r_1$ is the rate of the
  sub-code.)

  \indent(b) The replacement product of the two Cayley graphs is
  the Cayley graph \[C(A\rtimes B, S =(0,S_B)\cup \{ (a_i,0)
  |i=1,2,..,k\})\] on $N=2^p\cdot p$ vertices, where $A\rtimes B$
  is the semi-direct product group and the group operation is
  $(a,b)(c,d)=(a+\phi_b(c), b+d)$, for $a,c\in A,\ b,d\in B$.
  This is a regular graph with degree $d_g= k+|S_B|=k+2$. We
  interpret the vertices of the graph as sub-code constraints of
  a $[d_g,k_g,d_m]$ linear block code and the edges of the graph
  as code bits of the LDPC code, to obtain an LDPC code of block
  length $N_{LD}=2^p\cdot p\cdot d_g/2$ and rate
  \[
  r\ge \frac{N_{LD}-N(d_g-k_g)}{N_{LD}} =
  1-\frac{2(d_g-k_g)}{d_g}=\frac{2k_g}{d_g}-1.
  \]
  (Observe that $r\ge 2r_1-1$, where $r_1$ is the rate of the
  sub-code.) $\Box$ \label{ex1_zz}
\end{example}
\vspace{0.2in}

In some cases, to achieve a certain desired rate, we may have to
use a mixture of sub-code constraints from two or more linear
block codes. For example, to design a rate 1/2 LDPC code when
$d_g$ is odd, we may have to impose a combination of
$[d_g,k_g,d_{m1}]$ and $[d_g,k_g+1,d_{m2}]$ block code
constraints, for an appropriate $k_g$, on the vertices of the
graph.

\vspace{0.1in}

\begin{example} Let
  $B=SL_2(\mathbb{F}_p)$ be the group of all $2\times 2 $
  matrices over $\mathbb{F}_p$ with determinant one. Let {\small
    $ S_B =\left\{ \left(\begin{array}{cc}
          1&1\\
          0&1 \end{array}\right), \left(\begin{array}{cc}
          1&0\\
          1&1 \end{array}\right) \right\} $} be the generating
  set for the Cayley graph $C(B,S_B)$. Further, let
  $\mathbb{P}_1=\mathbb{F}_p\cup\{\infty\}$ be the projective
  line.  The {\em M$\ddot{o}$bius} action of $B$ on
  $\mathbb{P}_1$ is given
  by  $  \left(\begin{array}{cc} a&b\\ c&d \\
\end{array}\right)(x) = \frac{ax+b}{cx+d}$. Let $A
=\mathbb{F}_2^{\mathbb{P}_1}$ and let the action of $B$ on the
elements of $A$ be the M$\ddot{o}$bius permutation of the
coordinates as above. If we now choose $k$ elements
$a_1,a_2,\dots,a_k$ randomly from $A$ as in the previous example,
then {\rm \cite{al01p} again shows that with high probability,
  the Cayley graph
  $C(A,\{a_1^B,\dots,a_k^B\})$ is an expander.\\
  \indent(a) The zig-zag product of the two Cayley graphs is the
  Cayley graph \[C(A\rtimes
  B,S=\{(1_A,\beta)(a_i,1_B)(1_A,\beta')| \ \beta,\beta'\in S_B,
  i=1,2,..,k\})\] on $|A||B| = 2^{p+1}(p^3-p) $ vertices. (Note
  that $1_B=\left(\begin{array}{cc}
      1&0\\
      0&1 \end{array}\right)$ and $1_A= \left(\begin{array}{cc}
      0&0\\
      0&1 \end{array}\right)$.)  However, this Cayley graph will
  be a directed Cayley graph since the generating set $S$ is not
  symmetric. Hence, we modify our graph construction by taking
  two copies of the vertex set $A\rtimes B$. A vertex $v$ from
  one copy is connected to vertex $w$ in the other copy if there
  is a $s\in S$ such that $v*s = w$. The new product graph
  obtained has $2|A||B|$ vertices and every vertex has degree
  $d_g=|S|$; moreover, it is a balanced bipartite graph. An LDPC
  code of block length $|A||B|d_g$ is obtained by interpreting
  the vertices of the graph as sub-code constraints of a
  $[d_g,k_g,d_{m}]$ linear block code, and the edges as code bits
  of the LDPC code. The rate of this code is
\[r\ge 1-\frac{2(d_g-k_g)}{d_g} = \frac{2k_g}{d_g}-1.\]

\indent(b) The replacement product of the two Cayley graphs is
the Cayley graph \[C(A\rtimes B,S=(1_A,S_B)\cup
\{(a_i,1_B)|i=1,2,\dots,k\}),\] on $|A||B| = 2^{p+1}(p^3-p) $
vertices. Here also, the Cayley graph will be a directed Cayley
graph since the generating set $S$ is not symmetric. Hence, we
modify our graph construction by taking two copies of the vertex
set $A\rtimes B$. A vertex $v$ from one copy is connected to
vertex $w$ in the other copy if there is a $s\in S$ such that
$v*s = w$. The new product graph obtained has $2|A||B|$ vertices
and every vertex has degree $d_g=|S|$; moreover, it is a balanced
bipartite graph. An LDPC code of block length $|A||B|d_g$ is
obtained by interpreting the vertices of the graph as sub-code
constraints of a $[d_g,k_g,d_{m}]$ linear block code, and the
edges as code bits of the LDPC code. The rate of this code is
\[r\ge 1-\frac{2(d_g-k_g)}{d_g} = \frac{2k_g}{d_g}-1.\]
$\Box$}  \label{ex2_zz}
\end{example}\vspace{0.2in}

\begin{example}{\rm {\em Codes from unbalanced bipartite zig-zag product graphs}.\\
    Using a random construction, we design a $(c_1,d_1)$-regular
    bipartite graph $G_1$ on $(N,M)$ vertices. Similarly, we
    design a $(c_2,d_2)$-regular bipartite graph $G_2$ on
    $(d_1,c_1)$ vertices.  The zig-zag product of $G_1$ and $G_2$
    is a $(c_2^2,d_2^2)$-regular graph on $(N\cdot d_1,M\cdot
    c_1)$ vertices. An LDPC code is obtained as before by
    interpreting the degree $c_2^2$ vertices [resp. degree
    $d_2^2$ vertices] as sub-code constraints of a $C_{S1}
    =[c_2^2,k_1,d_{m1}]$ [resp. a $C_{S2}=[d_2^2,k_2,d_{m2}]$]
    linear block code and the edges of the product graph as code
    bits of the LDPC code. The block length of the LDPC code thus
    obtained is $N_{LD}=Nd_1c_2^2$ and the rate is \[r\ge
    \frac{Nd_1c_2^2-(Nd_1(c_2^2-k_1)+Mc_1(d_2^2-k_2))}{Nd_1c_2^2}
    = \frac{k_1}{c_2^2}+\frac{k_2}{d_2^2}-1\] (since $Nd_1c_2^2 =
    Mc_1d_2^2$ is the number of edges in the graph). Observe that
    $r \ge r_1+r_2-1$, where $r_1$ and $r_2$ are the rates of the
    two sub-codes $C_{S1}$ and $C_{S2}$, respectively. $\Box$}
  \label{ex3_zz}
\end{example}

\vspace{0.1in}

\section{Performance of Zig-zag and Replacement Product LDPC
  Codes}             \label{sec-6}

The performance of the LDPC code designs based on zig-zag and
replacement product graphs is examined for use over the additive
white Gaussian noise (AWGN) channel. (Binary modulation is
simulated and the bit error performance with respect to signal to
noise ratio (SNR) $E_b/N_o$ is determined.) The LDPC codes are
decoded using the graph based iterative sum-product (SP)
algorithm. Since LDPC codes based on product graphs use sub-code
constraints, the decoding at the constraint nodes is accomplished
using the BCJR algorithm on a trellis representation of the
appropriate sub-code. (A simple procedure to obtain the trellis
representation of the sub-code based on its parity check matrix
representation is discussed in~\cite{wo78}.) It must be noted
that as the number of states in the trellis representation and
the block length of the sub-code increases, the decoding
complexity correspondingly increases.

Figure~\ref{zigzag1} shows the performance of the zig-zag product
LDPC codes based on Example 4.1, with sum-product decoding. For
the parameters $p=5$ and $k=5$, five elements in
$A=\mathbb{F}_2^p$ are chosen (randomly) to yield a set of
generators for the Cayley graph of the semi-direct product group.
The Cayley graph has 160 vertices, each of degree 20. The
sub-code used for the zig-zag LDPC code design is a $[20,15,4]$
code and the resulting LDPC code has rate 1/2 and block length
1600. The figure also shows the performance of a LDPC code based
on a randomly designed degree 20 regular graph on 160 vertices
which also uses the same sub-code constraints as the former code.
The two codes perform comparably, indicating that the expansion
of the zig-zag product code compares well with that of a random
graph of similar size and degree. Also shown in the figure is the
performance of a $(3,6)$ regular LDPC code, that uses no special
sub-code constraints other than simple parity check constraints,
having the same block length and rate. Clearly, using strong
sub-code constraints improves the performance significantly,
albeit at the cost of higher decoding complexity. The figure also
shows another set of curves for a longer block length design.
Choosing $p=11$ and $k=5$ and the $[20,15,4]$ sub-code
constraints yields a rate 1/2 and block length 225,280 zig-zag
product LDPC code. At this block length also, the LDPC based on
the zig-zag product graph is found to perform comparably, if not,
better than the LDPC code based on a random degree 20 graph. The
zig-zag product graph has a poor girth\footnote{Note that there
  is no growth in the girth of the zig-zag product graph as
  opposed to that for a randomly chosen graph, with increasing
  graph size.} and this causes the performance of the zigzag LDPC
code to be inferior to that of the random LDPC codes at high
signal to noise ratios.

Figure~\ref{rep1} shows the performance of a replacement product
LDPC code based on Example 4.1, with sum-product decoding. For
the parameters $p=11$ and $k=13$, 13 elements in
$A=\mathbb{F}_2^p$ are chosen (randomly) to yield a set of
generators for the Cayley graph of the semi-direct product group.
The Cayley graph has 22,528 vertices, each of degree 15. The
sub-code used for the replacement product LDPC code design is a
$[15,11,3]$ Hamming code and the resulting LDPC code has rate
0.4667 and block length 168,960. The figure also shows the
performance of a LDPC code based on a randomly designed degree 15
regular graph on 22,528 vertices which also uses the same
sub-code constraints as the former code. Here again, the two
codes perform comparably, indicating that the expansion of the
replacement product code compares well with that of a random
graph of similar size and degree.

{\begin{center}
\begin{figure}
  \centering {
    \resizebox{3.1in}{2.8in}{\includegraphics{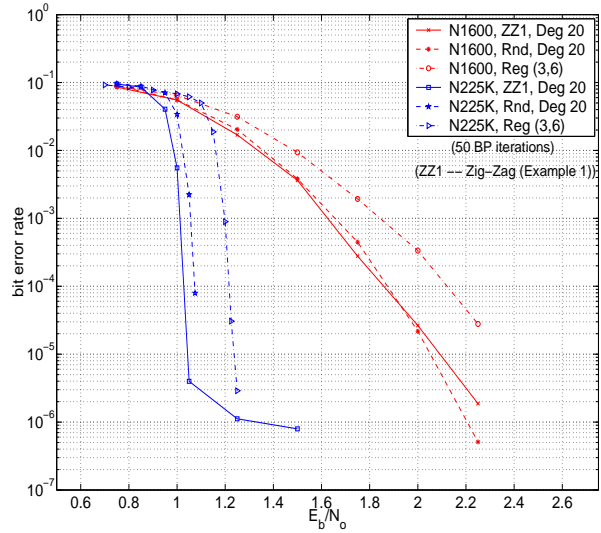}}}
             \caption{LDPC codes from zig-zag product graphs
               based on Example 4.1.}\label{zigzag1}
\end{figure}

\begin{figure}
  \centering {
    \resizebox{3.3in}{2.9in}{\includegraphics{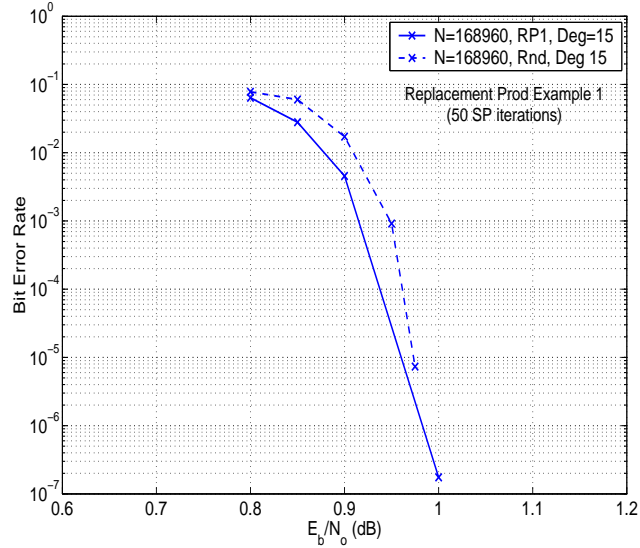}}}
             \caption{LDPC code from replacement product
               based on Example 4.1.}\label{rep1}
\end{figure}

\end{center}
}

Figure~\ref{zigzag2} shows the performance of zig-zag product
LDPC codes based on Example 4.2, with sum-product decoding. Once
again, this performance is compared with the analogous
performance of a LDPC code based on a random graph using
identical sub-code constraints and having the same block length
and rate. These results are also compared with a $(3,6)$ regular
LDPC code that uses simple parity check constraints.  For the
parameters $p=3$ and $k=5$ in Example 4.2, a bipartite graph,
based on the zig-zag product graph, on $768$ vertices with degree
20 is obtained. Using the $[20,15,4]$ sub-code constraints as
earlier, a block length 7680 rate 1/2 LDPC code is obtained. This
code performs comparably with the random LDPC code that is based
on a degree 20 randomly designed graph. Using the parameters
$p=5$ and $k=4$ and a $[16,12,2]$ sub-code, a longer block length
122,880 LDPC code is obtained. As in the previous case, this code
also performs comparably, if not, better than its random
counterpart for low to medium signal-to-noise ratios (SNRs). Once
again, we attribute its slightly inferior performance at high
SNRs to the poor girth of the zig-zag product graph.

Figure~\ref{rep2} shows the performance of a replacement product
LDPC code based on Example 4.2, with sum-product decoding. Once
again, this performance is compared with the analogous
performance of an LDPC code based on a random graph using
identical sub-code constraints and having the same block length
and rate. For the parameters $p=5$ and $k=13$ in Example 4.2, a
bipartite graph, based on the replacement product graph, on
$15,360$ vertices with degree 15 is obtained. Using the
$[15,11,3]$ Hamming code as a sub-code in the replacement product
graph, a block length 115,200 rate 0.4667 LDPC code is obtained.
The performance of the replacement product LDPC code is inferior
to that of the random code in this example due to the poor choice
of the generators in the component Cayley graphs.  We believe a
more judicious choice would improve the performance considerably.

{\begin{center}
\begin{figure}
  \centering{
    \resizebox{3.1in}{2.8in}{\includegraphics{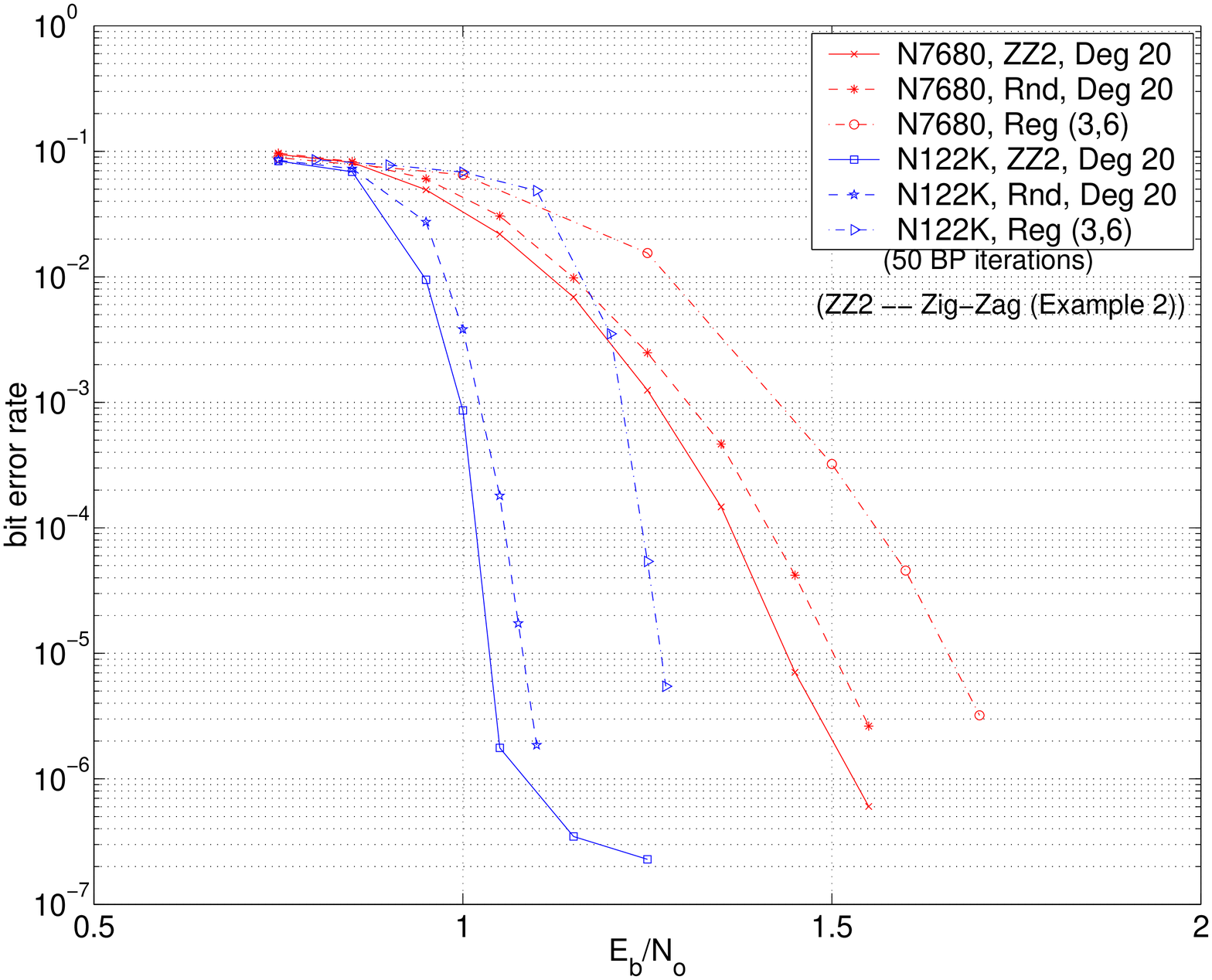}}}
            \caption{LDPC codes from zig-zag product graphs based on
              Example 4.2.}
     \label{zigzag2}
\end{figure}

\begin{figure}
  \centering {
    \resizebox{3.3in}{2.9in}{\includegraphics{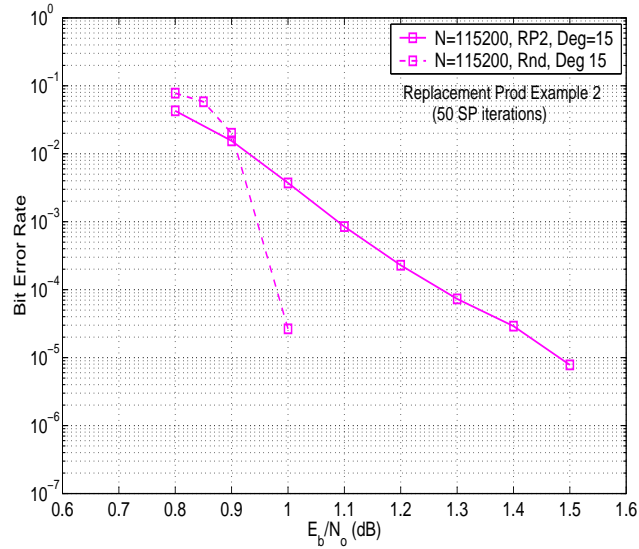}}}
             \caption{LDPC code from replacement product
               based on Example 4.2.}\label{rep2}
\end{figure}
\end{center}
}

Figure~\ref{zigzag3} shows the performance of LDPC codes designed
based on the zig-zag product of two unbalanced bipartite graphs
as in Example 4.3. A $(6,10)$-regular bipartite graph on
$(20,12)$ vertices is chosen as one of the component graphs and a
$(3,5)$-regular bipartite graph on $(10,6)$ vertices is chosen as
the other component. Their zig-zag product is a $(9,25)$-regular
bipartite graph on $(200,72)$ vertices. Using sub-code
constraints of two codes -- a $[9,6,2]$ and a $[25,21,2]$ linear
block code -- a block length 1800 LDPC code of rate 0.5066 is
obtained. The performance of this code is compared with a LDPC
code based on a random $(9,25)$-regular bipartite graph using the
same sub-code constraints, and also with a block length 1800
random $(3,6)$ regular LDPC code. All three codes perform
comparably, with the random $(3,6)$ showing a small improvement
over others at high SNRs.  Given that the zigzag product graph is
composed of two very small graphs, this result highlights the
fact that good graphs may be designed using just simple component
graphs.

\begin{figure}

  \centering{\resizebox{3.1in}{2.8in}{\includegraphics{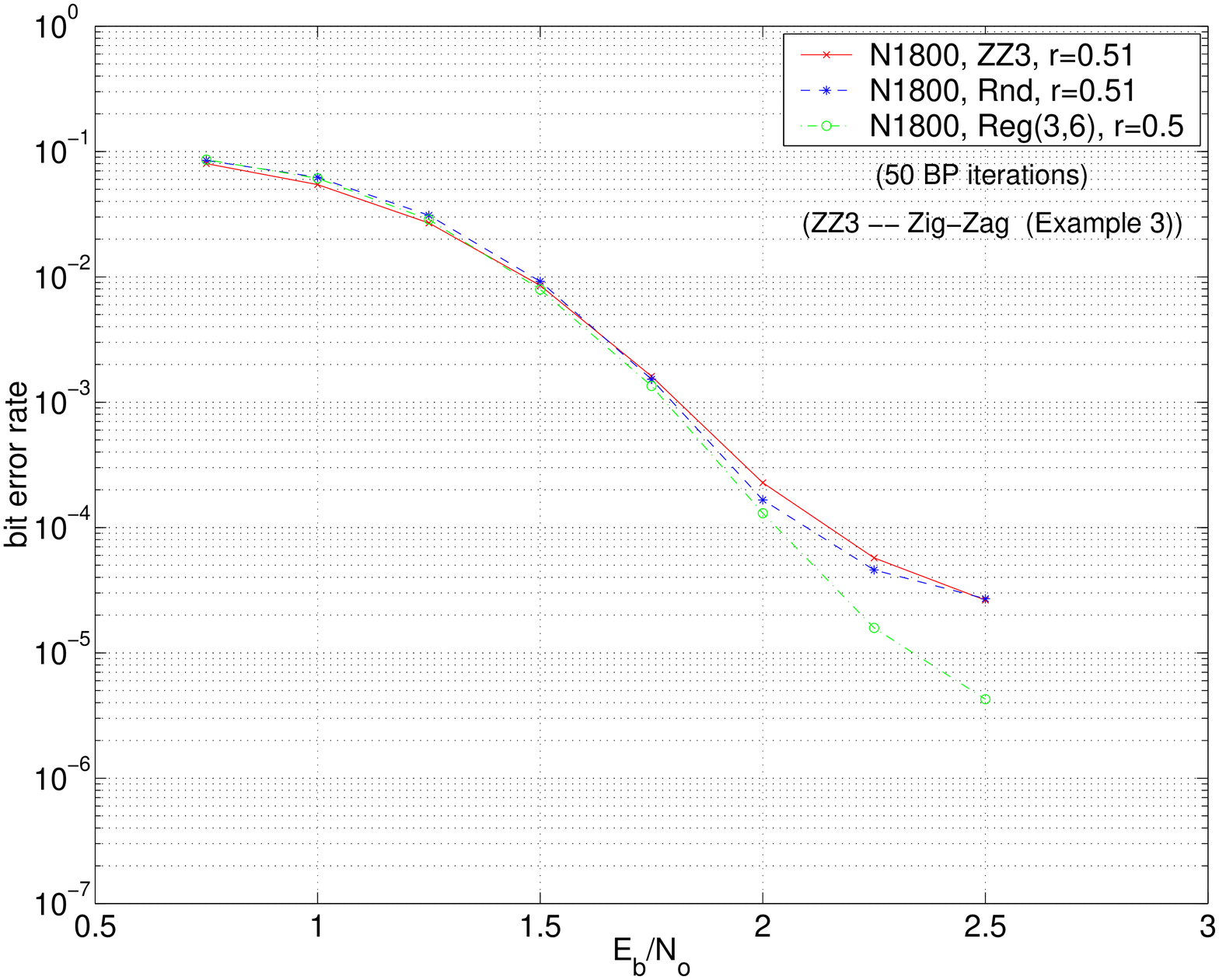}}}
\caption{LDPC code from the unbalanced bipartite zig-zag product
  graph based on Example 4.3.} \label{zigzag3}
\end{figure}

\vspace{0.1in}

\section{Iterative construction of generalized product graphs}
\label{sec-7}

In this section, we introduce iterative families of expanders
that address an important design problem in graph theory and that
have several other practical engineering applications such as in
designing communication networks, complexity theory, and
derandomization techniques.

For code constructions, we would ideally use products that could
be iterated to generate families of LDPC codes having a slow
growth in the number of vertices (so as to get codes for many
block-lengths), while maintaining a constant (small) degree. The
iterative families described in this section have these
characteristics, but unfortunately do not have parameters that
make the codes practical. Designing such iterative constructions
suitable for coding is a nice open problem.

First we review the iteration scheme of \cite{re02} for the
original zig-zag product starting from a seed graph $H$. The
existence of the seed graph $H$ as well as explicit examples of
suitable seed graphs for $H$ are also discussed in \cite{re02}.
We present new iterative constructions of a modified unbalanced
bipartite zig-zag product and the replacement product thereafter.
\vspace{0.1in}

\subsection{Iterative construction of original zig-zag product graphs}
We will need a squaring operation and the zig-zag operation in
the iterative technique that is proposed next.  Note that for a
graph $G$, its square $G^2$ is a graph whose vertices are the
same as in $G$ and whose edges are paths of length two in $G$.
Further, if $G$ is a $(N,D,\lambda)$ graph, then $G^2$ is a
$(N,D^2,\lambda^2)$ graph.

A graph $H$ is used to serve as the basic building block for the
iteration. Let $H$ be any $(D^4,D,\frac{1}{5})$ graph.  Then the
iteration is defined by
\[ G_1 = H^2 = (D^4,D^2,\frac{1}{25}).\]
\[G_{i+1}=G_i^2 \circz H.\]

The above iterative construction indeed gives a family of
expanders as presented in the following result: \vspace{0.1in}
\begin{theorem}\cite{re02}
  For every $i$, $G_i$ is an $(D^{4i},D^2,\frac{2}{5})$ graph.
\end{theorem}
\vspace{0.1in}

\subsection{Iterative construction of unbalanced bipartite zig-zag product graphs}
The unbalanced bipartite zig-zag product presented in
Section~\ref{sec-4} cannot be used directly to obtain an
iterative construction, due to constraints on the
parameters\footnote{ The only parameters that were compatible
  were for the special case where the bipartite components were
  balanced. }. Therefore, we slightly modify the zig-zag product
by introducing an additional step on the small component graph in
the product construction. We note that the introduction of this
additional step can only increase the expansion of the zig-zag
product graph.  However, this increase in expansion is at the
cost of increasing the degree of the graph slightly. The new
modified unbalanced bipartite zig-zag product is presented next,
followed by an iterative construction that uses this product.
\vspace{0.1in}

\subsubsection{Modified unbalanced bipartite zig-zag product}
The two component graphs are unbalanced bipartite graphs, i.e.,
the two sets of vertices have different degrees. Let $G_1$ be a
$(c_1,d_1)$-regular graph on the vertex sets $V_1, W_1$, where
$|V_1| = N$ and $|W_1| = M$. Let $G_2$ be a $(c_2,d_2)$-regular
graph on the vertex sets $V_2, W_2$, where $|V_2| = d_1$ and
$|W_2| = c_1$. Let $\lambda_{1}$ and $\lambda_{2}$ denote the
second largest eigenvalues of the normalized adjacency matrices
of $G_1$ and $G_2$, respectively. Again, randomly number the
edges around each vertex $\tilde{v} $ in $G_1$ and $G_2$ by
$\{1,\ldots, deg(\tilde{v})\}$, where $deg(\tilde{v})$ is the
degree of $\tilde{v}$. Then the zig-zag product graph, which we
will denote by $G=G_1\circzm G_{2}$, is a
$(c_2^2d_2,d_2^2c_2)$-regular bipartite graph on the vertex sets
$V, W$ with $|V|=N\cdot d_1$, $|W|=M\cdot d_1$, formed in the
following manner: \vspace{0.1in}

\begin{itemize}
\item Every vertex $v \in V_1$ and $w\in W_1$ of $G_1$ is
  replaced by a copy of $G_2$. The cloud at a vertex $v \in V_1$
  has vertices $V_2$ on the left and vertices $W_2$ on the right,
  with each vertex from $W_2$ corresponding to an edge from $v$
  in $G_1$. The cloud at a vertex $w\in W_1$ is similarly
  structured with each vertex in $V_2$ in the cloud corresponding
  to an edge of $w$ in $G_1$. (See
  Figure~\ref{zigzag_bipartiteU1}.)  Then the vertices from $V$
  are represented as ordered pairs $(v,k)$, for $v\in
  \{1,\dots,N\}$ and $k\in \{1,\dots,d_1\}$, and the vertices
  from $W$ are represented as ordered pairs $(w,\ell)$, for $w\in
  \{1,\dots,M\}$ and $\ell\in \{1,\dots,c_1\}$.

\item A vertex $(v,k) \in V$ is connected to a vertex in $W$ by
  making four steps in the product graph. The first three steps
  are the same as in Section~\ref{sec-4}. The fourth step is:

  \begin{itemize}

  \item A second small step from right to left in the local copy
    of $G_{2}$. This is a step $(v[k[i]],\ell[j]) \rightarrow
    (v[k[i]],\ell[j][j'])$, where the final vertex is in $W$, for
    $j' \in \{1,\ldots,d_2\}$.
   \end{itemize}

   Therefore, there is an edge between $(v,k)$ and
   $(v[k[i]],\ell[j][j'])$.
\end{itemize}
\vspace{0.1in}

\begin{theorem}
  Let $G_1$ be a $(c_1,d_1)$-regular bipartite graph on $(N, M)$
  vertices with $\lambda(G_1)=\lambda_1$, and let $G_2$ be a
  $(c_2,d_2)$-regular bipartite graph on $(d_1, c_1)$ vertices
  with $\lambda(G_2)=\lambda_2$. Then, the modified zig-zag
  product graph $G_1\circzm G_2$ is a
  $(c_2^2d_2,c_2d_2^2)$-regular bipartite on $(N\cdot d_1, M\cdot
  d_1)$ vertices with $\lambda=\lambda(G_1\circzm G_2) \le
  \lambda_1+\lambda_2+\lambda_2^2$.  Moreover, if $\lambda_1<1$
  and $\lambda_2<1$, then $\lambda=\lambda(G_1\circzm G_2) < 1$.
\label{zzmod_theorem}
\end{theorem}
\vspace{0.1in}

The proof is omitted but may be seen intuitively given the
expansion of the original unbalanced zig-zag product
(Theorem~\ref{zztheorem}) in the following way. The new step is
independent of the previous steps and is essentially a random
step on an expander graph ($G_2$). Considering a distribution on
the vertices $(v,k)$ of $G=G_1\circzm G_2$, if the distribution
of $k$ conditioned on $v$ is close to uniform after step 3, then
step 4 is redundant and no gain is made. If the distribution of
$k$ conditioned on $v$ is not close to uniform after step 3, then
step 4 will increase the entropy of $k$ by the expansion of
$G_2$.

\vspace{0.1in}

\subsubsection{Iterative construction}

The modified zigzag product of $G_1=(N,M, c_1,d_1,\lambda_1)$ and
$G_2=(d_1,c_1,c_2,d_2,\lambda_2)$ is a graph $G=G_1\circzm G_{2}$
that is $(c_2^2d_2, c_2d_2^2)$-regular on $(Nd_1,Md_1)$ vertices.
For the iteration, let $H$ be any $(N=c_2^4d_2^5,M=c_2^5d_2^4,
c_2, d_2, \lambda)$ expander graph, Then, the iteration is
defined by
\[ G_1=H^3=(N,M, c_2^2d_2,c_2d_2^2,\lambda^3).\]
\[ G_{i+1}=G_i^3 \circzm H.\]

We show that the above iterative technique yields a family of
expanders in the following \vspace{0.1in}

\begin{theorem}
  Let $H$ be a $(c_2d_2^2,c_2^2d_2,c_2,d_2,\lambda)$ graph, where
  $\lambda \le 0.296$. Let $G_1=H^3$ and $G_{i+1}=G_i^3 \circzm
  H$. Then the $i$-th iterated zig-zag product graph $G_i$ is a
  $((c_2^{4i}d_2^{5i},c_2^{4i+1}d_2^{5i-1},c_2^2d_2,c_2d_2^2,
  \lambda')$ graph, where $\lambda' \le 0.55$.
\label{zzmod_iteration_thm}
\end{theorem}
\vspace{0.1in}

\begin{proof}
  Let $n_i$ and $m_i$ be the number of left vertices and right
  vertices in $G_i$, respectively.  Since $G_i =G_{i-1}^3 \circzm
  H$, we have $n_i = n_{i-1}(c_2^4d_2^5)$ and $m_i =
  m_{i-1}(c_2^4d_2^5)$. Since $n_1 = c_2^4d_2^5$ and $m_1 =
  c_2^5d_2^4$, it follows from the above recursion that $n_i =
  c_2^{4i}d_2^{5i}$ and $m_i = c_2^{4i+1}d_2^{5i-1}$. Note that
  $G_i$ is always $(c_2^2d_2, c_2d_2^2)$-regular.

  Let $\lambda_i$ be the normalized second eigenvalue of $G_i$.
  Using the result from Theorem~\ref{zzmod_theorem}, we have
\[ \lambda_i \le \lambda_{i-1}^3 + \lambda + \lambda^2.\] Further
note that $\lambda_1 = \lambda^3$.  Observe that even for
$\lambda_i = \lambda_{i-1}^3 + \lambda + \lambda^2$, the series
converges $\lambda_i \rightarrow 0.5499$ when $\lambda \le
0.296$.  Hence, for each iteration $i$, $\lambda_i \le 0.55$,
thereby yielding a family of expanders.
\end{proof}
\vspace{0.1in}

\subsection{Iterative construction of replacement product graphs}

The replacement product of $G_1=(N,d_1,\lambda_1)$ and
$G_2=(d_1,d_2,\lambda_2)$, denoted by $G_1 \circr G_2$, is an
$(Nd_1, d_2+1, \lambda)$ graph. In \cite{re02}, it is shown that
the expansion of the replacement product graph is given by
\begin{eqnarray}
\lambda \le (p+(1-p)f(\lambda_1, \lambda_2))^{\frac{1}{3}},
\hspace{1.5in}
\label{eq1}\\
\mbox{ where } p = \frac{d_2^2}{(d_2+1)^3} \mbox{  and  } f(\lambda_1, \lambda_2)
=
\frac{1}{2}(1-\lambda_2^2)\lambda_1+\frac{1}{2}\sqrt{(1-\lambda_2^2)^2
\lambda_1^2+4\lambda_2^2}.\nonumber
\end{eqnarray}
To obtain an iterative construction, we choose two graphs $G_1 =
(N,(d+1),\lambda_1)$ and $H=((d+1)^4,d,\lambda_2)$.

The iteration is defined by
\[ G_{i+1}=(G_i)^4 \circr H.\]

We show that the above iterative construction results in a family
of expanders. \vspace{0.1in}

\begin{theorem}
  Let $G_1$ be a $(N,(d+1),\lambda_1)$ graph and let $H$ be a
  $((d+1)^4,d,\lambda_2)$ graph, where $\lambda_1 \le 0.2,
  \lambda_2 \le 0.2$ and $d \ge 6$. Let $G_{i+1}=(G_i)^4 \circr
  H$.  Then the $i$-th iterated replacement product graph $G_i$
  is a $(N(d+1)^{4(i-1)},d+1,\lambda)$ graph, where $\lambda \le
  0.86$.
\label{rep_iteration_thm}
\end{theorem}
\vspace{0.1in}

\begin{proof}
  Let $n_i$ be the number of vertices in $G_i$. Then $n_i =
  n_{i-1}(d+1)^4$. Since $n_1 = N$, it follows that $n_i =
  N(d+1)^{4i-4}$. It is clear that the degree of $G_i$ is one
  more than the degree of $H$, and thus, $G_i$ is
  $(d+1)$-regular. Let $\lambda_i$ be the normalized second
  eigenvalue of $G_i$. Using the result from
  Equation~(\ref{eq1}), we have \[ \lambda_i \le
  [p+(1-p)f(\lambda_{i-1}^4, \lambda_2)]^{\frac{1}{3}},\] where
  $p = \frac{d^2}{(d+1)^3}$ and $f$ is as above. Using numerical
  methods with Matlab, it was verified that $\lambda_i$ converges
  to $0.8574$ when $\lambda_1 \le 0.2, \lambda_2 \le 0.2$, and $d
  \ge 6$. Hence, for each iteration $i$, $\lambda_i < 0.86$,
  thereby yielding a family of expanders.
\end{proof}
\vspace{0.1in}

Note that if the above iteration was defined to be $
G_{i+1}=(G_i)^2 \circr H$, then for no choice of $\lambda_1,
\lambda_2$, or $d$, would the resulting iterative family be
expanders. \vspace{0.1in}

\section{Conclusions}             \label{sec-8}

In this paper we generalized the zig-zag product resulting in a
product for unbalanced bipartite graphs. We proved that the
resulting graphs are expander graphs as long as the component graphs
are expanders.  We examined the performance of LDPC codes obtained
from zig-zag and replacement product graphs. The resulting product
LDPC codes perform comparably to random LDPC codes with the
additional advantage of having a compact description. We also
introduced iterative constructions for the unbalanced zig-zag and
replacement products yielding families of graphs with small,
constant or slowly increasing degrees, and good expansion. Although
these iterative schemes do not yield parameters for practical codes
as yet, we believe they provide a first step in exploring iterated
products for code construction.  Designing iterative graph products
that result in a family of good practical codes remains an
intriguing open problem. We conclude that codes from product graphs
provide a nice avenue for code constructions.

\vspace{0.1in}


\begin{thebibliography}{10}

\bibitem{al86} N.~Alon.  \newblock Eigenvalues and expanders.
  \newblock {\em Combinatorica}, 6(2):83--96, 1986.  \newblock
  Theory of computing (Singer Island, Fla., 1984).

\bibitem{al01p} N.~Alon, A.~Lubotzky, and A.~Wigderson.
  \newblock Semi-direct product in groups and zig-zag product in
  graphs: connections and applications (extended abstract).
  \newblock In {\em 42nd IEEE Symposium on Foundations of
    Computer Science (Las Vegas, NV, 2001)}, pages 630--637. IEEE
  Computer Soc., Los Alamitos, CA, 2001.

\bibitem{im00} W.~Imrich and S.~Klav{\v{z}}ar.  \newblock {\em
    Product graphs}.  \newblock Wiley-Interscience Series in
  Discrete Mathematics and Optimization.  Wiley-Interscience, New
  York, 2000.  \newblock Structure and recognition, With a
  foreword by Peter Winkler.

\bibitem{ja03} H.~Janwa and A.~K. Lal.  \newblock On {T}anner
  codes: minimum distance and decoding.  \newblock {\em Appl.
    Algebra Engrg. Comm. Comput.}, 13(5):335--347, 2003.

\bibitem{ke03p} C.~Kelley, J.~Rosenthal, and D.~Sridhara.
  \newblock Some new algebraic constructions of codes from graphs
  which are good expanders.  \newblock In {\em Proc. of the 41-st
    Allerton Conference on Communication, Control, and
    Computing}, pages 1280--1289, 2003.

\bibitem{ke06u1} C.~Kelley and D.~Sridhara.  \newblock Eigenvalue
  bounds on the pseudocodeword weight of expander codes.
  \newblock {\em Journal of Advances in Mathematics of
    Communication}, Aug. 2007.

\bibitem{la00p} J.~Lafferty and D.~Rockmore.  \newblock Codes and
  iterative decoding on algebraic expander graphs.  \newblock In
  the Proceedings of ISITA 2000, Honolulu, Hawaii, available at
  \href{http://www-2.cs.cmu.edu/afs/cs.cmu.edu/user/lafferty/www/pubs.html}%
{http://www-2.cs.cmu.edu/afs/cs.cmu.edu/user/lafferty/www/pubs.html},
  November 2000.

\bibitem{li03r} N.~Linial and A.~Wigderson.  \newblock Expander
  graphs and their applications.  \newblock Lecture notes of a
  course given at the Hebrew University, 2003.
\newblock Available under \href{http://www.math.ias.edu/~avi/TALKS/}%
{http://www.math.ias.edu/$\sim$avi/TALKS/}.

\bibitem{lu94} A.~Lubotzky.  \newblock {\em Discrete Groups,
    Expanding Graphs and Invariant Measures}.  \newblock
  Birkh\"auser Verlag, Basel, 1994.  \newblock With an appendix
  by Jonathan D. Rogawski.

\bibitem{lu88a} A.~Lubotzky, R.~Phillips, and P.~Sarnak.
  \newblock {R}amanujan graphs.  \newblock {\em Combinatorica},
  8(3):261--277, 1988.

\bibitem{ma88a2} G.~A. Margulis.  \newblock Explicit
  group-theoretic constructions of combinatorial schemes and
  their applications in the construction of expanders and
  concentrators.  \newblock {\em Problems Inform. Transmission},
  24(1):39--46, 1988.  \newblock Translation from Problemy
  Peredachi Informatsii.

\bibitem{me03} R.~Meshulam and A.~Wigderson.  \newblock Expanders
  in group algebras.  \newblock {\em Combinatorica}, 2003.
  \newblock To appear.

\bibitem{re02} O.~Reingold, S.~Vadhan, and A.~Wigderson.
  \newblock Entropy waves, the zig-zag graph product, and new
  constant-degree expanders.  \newblock {\em Ann. of Math. (2)},
  155(1):157--187, 2002.

\bibitem{ro00p} J.~Rosenthal and P.~O. Vontobel.  \newblock
  Constructions of {LDPC} codes using {R}amanujan graphs and
  ideas from {M}argulis.  \newblock In {\em Proc. of the 38-th
    Allerton Conference on Communication, Control, and
    Computing}, pages 248--257, 2000.

\bibitem{si96} M.~Sipser and D.~A. Spielman.  \newblock Expander
  codes.  \newblock {\em IEEE Trans. Inform. Theory}, 42(6, part
  1):1710--1722, 1996.

\bibitem{ta81a} R.~M. Tanner.  \newblock A recursive approach to
  low complexity codes.  \newblock {\em IEEE Trans. Inform.
    Theory}, 27(5):533--547, 1981.

\bibitem{ta84} R.~M. Tanner.  \newblock Explicit concentrators
  from generalized ${N}$-gons.  \newblock {\em SIAM J. Algebraic
    Discrete Methods}, 5(3):287--293, 1984.

\bibitem{ti97} J.-P. Tillich and G.~Z{\'e}mor.  \newblock Optimal
  cycle codes constructed from {R}amanujan graphs.  \newblock
  {\em SIAM J. Discrete Math.}, 10(3):447--459, 1997.

\bibitem{wo78} J.~K. Wolf.  \newblock Efficient maximum
  likelihood decoding of linear block codes using a trellis.
  \newblock {\em IEEE Trans. Inform. Theory}, IT-24(1):76--80,
  1978.
\end{thebibliography}
\def\cprime{$'$} \def\cprime{$'$}

\medskip

Submitted: August 17, 2007.

\medskip

\end{document}